\documentclass[11pt,twoside]{article}
\usepackage{atmp04}
\copyrightnotice{2004}{8}{83}{139}   

\usepackage{latexsym}
{\catcode `\@=11 \global\let\AddToReset=\@addtoreset}
\AddToReset{equation}{section}
\renewcommand{\theequation}{\thesection.\arabic{equation}}

\usepackage{cite}
\usepackage{url}
\usepackage{amsfonts}
\usepackage{a4}

\newcommand{\beaa}{\begin{eqnarray*}}
\newcommand{\eeaa}{\end{eqnarray*}}

\newcommand{\btw}{{}^2{b}}

\newcommand{\dtwo}{\triangle}
\newcommand{\Gst}{\Gamma}
\newcommand{\three}{x}
\newcommand{\myzero}{u}

\newcommand{\funk}{\mathcal{F}}
\newcommand{\Der}{\mathcal{D}}
\newcommand{\real}{\mathbb {R}}

\newcommand{\uder}{\wr} 

\newcommand{\gtw}{{}^2{g}}

\newcommand{\Gtrzy}{{}^3{\Gamma}}

\newcommand{\gmetric}{g} 

\newcommand{\zk}{{\mathring k}}
\newcommand{\zlambda}{{\mathring \lambda}}

\newcommand{\mcG}{{\mycal G}}%
\newcommand{\const}{\mbox{\rm const}}
\newcommand{\rd}{{\rm
d}}\newcommand{\newW}{u}

\newcommand{\mtb}{m_{\rm \scriptsize TB}}
\newcommand{\tr}{\mathrm{tr}\,}
\newcommand{\trz}{\mathrm{tr}_{b_0}\,}
\newcommand{\id}{\mathrm{id}}

\newcommand{\znabla}{\mathring{\nabla}}
\newcommand{\zGamma}{\mathring{\Gamma}}
\newcommand{\zn}{\mathring{n}}
\newcommand{\hst}{{\breve{h}}} 
\newcommand{\thb}{b}
\newcommand{\thg}{g}
\newcommand{\zomega}{{\mathring\omega}}
\newcommand{\fourg}{{}^4g}
\newcommand{\bfourg}{{}^4\tg}
\newcommand{\fourb}{{}^4b}
\newcommand{\spt}{(\mcM,\fourg)}
\newcommand{\trK}{\mathrm{tr}_gK}
\newcommand{\Ol}{O_{\ln^*x }}
\newcommand{\tea}{\tilde e_a}
\newcommand{\teb}{\tilde e_b}

\newcommand{\bM}{\overline{M}}
\newcommand{\bmcM}{\,\,\,\,{\overline{\!\!\!\!\mcM}}}
\newcommand{\tM}{\bM}
\newcommand{\bfg}{^{4}{\overline{ g}}}
\newcommand{\tfg}{\bfg}

\newcommand{\repf}{{\hat f}}
\newcommand{\repe}{{\hat e}}

\newcommand{\tg}{{\overline{g}}}
\newcommand{\can}{g_{\cerc^{n-1}}}

\newcommand{\bel}[1]{\begin{equation}\label{#1}}

\newcommand{\mcM}{{\mycal M}}
\newcommand{\mcD}{{\mycal D}}

\newcommand{\lie}{{\mycal L}}

\DeclareFontFamily{OT1}{rsfs}{}
\DeclareFontShape{OT1}{rsfs}{m}{n}{ <-7> rsfs5 <7-10> rsfs7 <10->
rsfs10}{} \DeclareMathAlphabet{\mycal}{OT1}{rsfs}{m}{n}
\def\scri{{\mycal I}}%
\def\scrip{\scri^{+}}%
\def\Scri{\scri}

\newtheorem{defi}{\sc Definition\rm}[section]

\newtheorem{theorem}[defi]{\sc Theorem\rm}

\newtheorem{pro}[defi]{\sc Proposition\rm}

\newtheorem{definition}[defi]{\sc Definition\rm}

\newtheorem{theo}[defi]{\sc Theorem\rm}

\newtheorem{Lemma}[defi]{\sc Lemma\rm}

\newtheorem{remk}[defi]{\sc Remark\rm}
\newtheorem{Remark}[defi]{\sc Remark\rm}

\newcommand{\hyp}{{\mycal S}}
\newcommand{\bhyp}{\,\,\overline{\!\!{\mycal S}}}

\newcommand{\thyp}{\bhyp} 

\newcommand{\piM}{\partial_\infty M}
\newcommand{\piMx}{\partial_\infty \Mext}

\newcommand{\N}{{\mathbb N}}

\newcommand{\R}{{\mathbb R}}

\newcommand{\ra}{\rangle}
\newcommand{\la}{\langle}

\newcommand{\Eq}[1]{Equation~\eq{#1}}
\newcommand{\Eqsone}[1]{Equations~\eq{#1}}
\newcommand{\Eqs}[2]{Equations~\eq{#1}-\eq{#2}}

\newcommand{\trg }{\mathrm{tr}_g }
\newcommand{\trb }{\mathrm{tr}_b }
\newcommand{\cerc}{{\mathbb S}}

\newcommand{\Ric}{\mathrm{Ric}}

\newcommand{\dirac}{\mathfrak{D}}
\newcommand{\myqed }{\hfill $\Box$}

\newcommand{\Int}{\mathrm{int}}
\newcommand{\Mext}{M_{\mbox{{\scriptsize\rm ext}}}}
\newcommand{\zh}{\breve{h}}

\newcommand{\zN}{{\mathring N}}
\newcommand{\zD}{{\mathring D}}
\newcommand{\zK}{{\mathring K}}
\newcommand{\zA}{{\mathring A}}
\newcommand{\zS}{{\mathring S}}
\newcommand{\zB}{{\mathring B}}
\newcommand{\zP}{{\mathring P}}

\newcommand{\cNbz}{\mathcal{N}_{b_0,\zK_0}}

\newcommand{\ourU}{\mathbb U}
\newcommand{\ourV}{\mathbb V}

\newcommand{\be}{\begin{equation}}
\newcommand{\ee}{\end{equation}}
\newcommand{\Bgamma}{{B}} 
\newcommand{\bmetric}{{\fourb}} 

\newcommand{\Kp}{{\frak g}} 
\newcommand{\KA}{p} 
\newcommand{\ourW}{\mathbb W}

\newcounter{mnotecount}[section]
\renewcommand{\themnotecount}{\thesection.\arabic{mnotecount}}
\newcommand{\mnote}[1]
{\protect{\stepcounter{mnotecount}}$^{\mbox{\footnotesize  $
      \bullet$\themnotecount}}$ \marginpar{\raggedright\tiny
    $\!\!\!\!\!\!\,\bullet$\themnotecount: #1} }

\newcommand{\eq}[1]{(\ref{#1})}

\begin{document}
\arxurl{gr-qc/0307109}
\title{The Trautman-Bondi mass\\ of hyperboloidal initial data sets}
\author{Piotr T. Chru\'sciel$^{1,2}$,
Jacek Jezierski$^{1,3}$ and Szymon \L \c eski$^{1,4}$}
\address{$^1$Albert Einstein Institut f\"ur Gravitationsphysik\\
Golm, Germany}
\address{$^2$D\'epartement de Math\'ematiques, Parc de
Grandmont\\ F-37200 Tours, France}
\address{$^3$Katedra Metod Matematycznych Fizyki, Uniwersytet Warszawski\\
ul. Ho\.{z}a 74, 00-682 Warsaw, Poland}
\address{$^4$Centrum Fizyki Teoretycznej, Polska Akademia Nauk\\
Al. Lotnik\'{o}w 32/46, 02-668 Warsaw, Poland}
\addressemail{chrusciel@univ-tours.fr, Jacek.Jezierski@fuw.edu.pl, szleski@cft.edu.pl}

\markboth{\it The Trautman-Bondi mass}{\it P.T. Chru\'sciel, J.
Jezierski and Sz. \L\c eski}
\begin{abstract}
We give a definition of mass for conformally compactifiable
initial data sets.  The asymptotic conditions are compatible with
existence of gravitational radiation, and the compactifications
are allowed to be polyhomogeneous. We show that the resulting mass
is a geometric invariant, and we prove positivity thereof  in the
case of  a spherical conformal infinity. When $R(g)$ -- or,
equivalently, $\trg K$ -- tends to a negative constant to order
one at infinity, the definition is expressed purely in terms of
three-dimensional or two-dimensional objects.
\end{abstract}\cutpage
\tableofcontents

\section{Introduction}
\label{sec:Introdu}

 In 1958  Trautman~\cite{T}
(see also~\cite{Tlectures}) has introduced a notion of energy
suitable for asymptotically Minkowskian radiating gravitational
fields, and proved its decay properties; this mass has been
further studied by Bondi \emph{et al.}\/~\cite{BondiEtal62} and
Sachs~\cite{Sachs}. Several other definitions of mass have been
given in this setting, and to put our results in proper
perspective it is convenient to start with a general overview of
the subject. First, there are at least seven methods for defining
energy-momentum (``mass" for short) in the current
context:

\begin{enumerate}
\item A definition of Trautman~\cite{T}, based on the Freud integral~\cite{Freud39}, that involves
 asymptotically Minkowskian coordinates in space-time.  The
 definition stems from a Hamiltonian analysis in a fixed global
 coordinate system.

\item A definition of Bondi \emph{et al.}\/~\cite{BondiEtal62}, which uses space-time Bondi
coordinates.

\item A definition of Abbott-Deser~\cite{AbbottDeser}, originally introduced in the
context of space-times with negative cosmological constant, which
(as we will see) is closely related to the problem at hand. The
Abbott-Deser integrand turns out to coincide with the
linearisation of the Freud integrand, up to a total
divergence~\cite{ChNagyATMP}.

\item The space-time ``charge integrals",
derived in a geometric Hamiltonian
framework~\cite{ChAIHP,HT,ChNagy,CJK}. A conceptually distinct,
but closely related, variational approach, has been presented
in~\cite{LauYorkBrown}.

\item The ``initial data charge integrals", presented below,
expressed in terms of  data $(g,K)$ on an initial data manifold.

\item The Hawking mass and its variations such as the Brown-York mass, using
two-dimensional spheres.

\item A purely Riemannian definition, that provides a notion of  mass for asymptotically hyperbolic
Riemannian metrics~\cite{Wang,ChHerzlich}.
\end{enumerate}

Each of the above typically comes with several distinct
variations.

 Those definitions have the following properties:

\begin{enumerate}
\item The Bondi mass $m_B$ requires in principle a space-time on which Bondi
coordinates can be introduced. However, a null hypersurface
extending to future null infinity suffices. Neither of this is
directly adapted to an analysis in terms of usual spacelike
initial data sets. $m_B$ is an invariant under Bondi-van der
Burg-Metzner-Sachs coordinate transformations. Uniqueness is not
clear, because there could exist Bondi coordinates which are not
related to each other by a Bondi-van der Burg-Metzner-Sachs
coordinate transformation.

\item Trautman's definition $m_T$ requires existence of a certain class of asymptotically Minkowskian coordinates,
with $m_T$ being invariant  under a class of coordinate
transformations that arise naturally in this context~\cite{T}. The
definition is obtained by evaluating the Freud integral in
Trautman's coordinates.
    Asymptotically Minkowskian coordinates associated with the
    Bondi coordinates belong to the Trautman class, and the definition is invariant
    under a natural class of coordinate transformations. Trautman's conditions for existence of mass
    are less stringent, at least in principle\footnote{It is difficult to make a clear cut statement here because
existence theorems that lead to space-times with Trautman
coordinates seem to provide Bondi coordinates as well (though
perhaps in a form that is weaker than required in the original
definition of mass, but still compatible with an extension of
Bondi's definition).}, than the Bondi ones. $m_B$ equals $m_T$ of
the associated quasi-Minkowskian coordinate system, whenever $m_B$
exists as well. Uniqueness is not clear, because there could exist
Trautman coordinates which are not related to each other by the
coordinate transformations considered by Trautman.

\item The Hawking mass, and its variations, are \emph{a priori} highly sensitive to
the way that a family of spheres approaches a cut of $\Scri$.
(This is a major problem one faces when trying to generalise the
proof of the Penrose inequality to a hyperboloidal setting.) It is
known that those masses converge to the Trautman-Bondi mass when
evaluated on Bondi spheres, but this result is useless in a Cauchy
data context, as general initial data sets will not be collections
of Bondi spheres.

\item We will show below that the linearisation of the Freud integral coincides
with the linearisation of
the initial data charge integrals.

\item We will show below that the linearisation of the Freud integral does \emph{not}
coincide in general with the Freud integral. However, we will also
show that the resulting numbers coincide when decay conditions,
referred to as \emph{strong decay conditions}, are imposed. The
strong decay conditions turn out to be incompatible with existence
of gravitational radiation.

\item We will show below that a version of the Brown-York mass, as
well as the Hawking mass, evaluated on a specific foliation within
the initial data set, converges to the Trautman-Bondi mass.

\item We will show below that the Freud integrals coincide with the initial data charge
integrals, for asymptotically CMC initial data sets on which a
space-equivalent of Bondi coordinates can be constructed. The
proof is indirect and uses the result, just mentioned, concerning
the Brown-York mass.

\item It is important to keep in mind that hyperboloidal initial
data sets in general relativity arise in two different contexts:
as hyperboloidal hypersurfaces in asymptotically Minkowskian
space-times on which $K$ approaches a multiple of $g$ as one
recedes to infinity, or for spacelike hypersurfaces in space-times
with a negative cosmological constant on which $|K|_g$ approaches
zero as one recedes to infinity (compare~\cite{Kannar:adS}). This
indicates that the Abbott-Deser integrals, which arose in the
context of space-times with non-zero cosmological constant, could
be related to the Trautman-Bondi mass. It follows from our
analysis below that they do, in fact, coincide with the initial
data charge integrals under the strong decay conditions. In
space-times with a cosmological constant, the strong decay
conditions are satisfied on hypersurfaces which are, roughly
speaking, orthogonal to high order to the conformal boundary, but
will not be satisfied on more general hypersurfaces.

\item It has been shown in~\cite{ChHerzlich} that the strong decay
conditions on $g$ are \emph{necessary} for a well defined
Riemannian definition of mass and of momentum. This appears to be
paradoxical at first sight, since the strong decay conditions are
incompatible with gravitational radiation; on the other hand one
expects the Trautman-Bondi mass to be  well defined  even if there
is gravitational radiation. We will show below that the initial
data charge integrals  involve  delicate cancelations between $g$
and $K$, leading to a well defined notion of mass of initial data
sets without the stringent restrictions of the Riemannian
definition (which does not involve $K$).

\item Recall that for initial data sets which are asymptotically
flat in spacelike directions, one can define the mass purely in
terms of the induced three-dimensional metric. An unexpected
consequence of our analysis below is therefore that any definition
of mass of hyperboloidal initial data sets, compatible with
initial data sets containing gravitational radiations, must
involve the extrinsic curvature $K$ in a non-trivial way.

\end{enumerate}

 Let us discuss in some more detail  the
definition of the Trautman-Bondi mass. The
 standard current understanding of this
object is the following: one introduces a conformal completion at
null infinity of $(M,g)$ and assigns a mass $\mtb$ to sections of
the conformal boundary $\Scri$ using Bondi coordinates or
Newman-Penrose coefficients. The Trautman-Bondi mass $m_{\mbox{\rm
\scriptsize TB}}$ has been shown to be the unique functional,
within an appropriate class, which is non-increasing with respect
to deformations of the section to the future~\cite{CJM}.  A
formulation of that mass in terms of ``quasi-spherical" foliations
of null cones has been recently given in~\cite{Bartnik:qlnm} under
rather weak differentiability conditions.

The above approach raises some significant questions. First, there
is a potential ambiguity arising from the possibility of existence
of non-equivalent conformal completions of a Lorentzian space-time
(see~\cite{Chrusciel:2002mi} for an explicit example). From a
physical point of view, sections of Scri represent the asymptotic
properties of a radiating system at a given moment of retarded
time. Thus, one faces the curious possibility that two different
masses could be assigned to the same state of the system, at the
same retarded time, depending upon which of the conformally
inequivalent completions one chooses. While there exist some
partial results in the literature concerning the equivalence
question~\cite{Schmidtcqg91,GerochEspositoWitten}, no uniqueness
proof suitable for the problem at hand can be found. The first
main purpose of this paper is to show that the Trautman-Bondi mass
of a section $S$ of $\scrip$ is a geometric invariant in a sense
which is made precise in Section~\ref{Sginv} below.

Next, recall that several arguments aiming to prove positivity of
$\mtb$ have been given ({\em cf.}\/, amongst
others,~\cite{ReulaTod,HorowitzTod,LudvigsenVickers83b,schoen:yau:bondi}).
However, the published proofs are either incomplete or wrong, or
prove positivity of something which might be different from the TB
mass, or are not detailed enough to be able to form an opinion.
The second main purpose of this work is to give a complete proof
of positivity of $m_{\mbox{\rm \scriptsize TB}}$, see
Theorems~\ref{Tpos} and~\ref{Tposvac}.

An interesting property of the Trautman-Bondi mass is that it can
be given a Hamiltonian interpretation~\cite{CJK}. Now, from a
Hamiltonian point of view, it is natural to assign a Hamiltonian
to an initial data set $(\hyp,g,K)$, where $\hyp$ is a
three-dimensional manifold, without the need of invoking a
four-dimensional space-time. If there is an associated conformally
completed space-time in which the completion $\bhyp$ of $\hyp$
meets $\scrip$ in a sufficiently regular, say differentiable,
section $S$, then $\mtb(\hyp)$ can be defined as the
Trautman-Bondi mass of the section $S$. (Hypersurfaces satisfying
the above are called hyperboloidal.) It is, however, desirable to
have a definition which does not involve any space-time
constructions. For example, for non-vacuum initial data sets an
existence theorem for an associated space-time might be lacking.
Further, the initial data might not be sufficiently differentiable
to obtain an associated space-time. Next, there might be loss of
differentiability during evolution which will not allow one to
perform the space-time constructions needed for the space-time
definition of mass. Finally, a proof of uniqueness of the
definition of mass of an initial data set could perhaps be easier
to achieve than the space-time one. Last but not least, most
proofs of positivity  use three-dimensional hypersurfaces anyway.
For all those reasons it seems of interest to obtain a definition
of mass, momentum, \emph{etc.}, in an initial data setting. The
third main purpose of this work is to present such a definition,
see \eq{mi} below.

The final  $(3+1)$-dimensional formulae for the Hamiltonian
charges turn out to be rather complicated. We close this paper by
deriving a considerably simpler expression for the charges in
terms of the geometry of ``approximate Bondi spheres" near
$\scrip$, \Eqs{int2}{int2.1} below. The expression is similar in
spirit to that of Hawking and of Brown, Lau and
York~\cite{LauYorkBrown}. It applies to mass as well as momentum,
angular momentum and centre of mass.

It should be said that our three-dimensional and two-dimensional
versions of the  definitions do not cover all possible
hyperboloidal initial hypersurfaces, because of a restrictive
assumption on the asymptotic  behavior of $g^{ij}K_{ij}$, see
\eq{Kres}.  This condition arises from the need to reduce the
calculational complexity of our problem; without \eq{Kres} we
would probably not have been able to do all the calculations
involved. We do not expect this condition to be essential, and we
are planning to attempt to remove it using computer algebra in the
future; the current calculations seem pretty much to be the limit
of what one can calculate by hand with a reasonable degree of
confidence in the final formulae. However, the results obtained
are sufficient to prove positivity of $\mtb(S)$ for all smooth
sections of $\scrip$ which bound some smooth complete hypersurface
$\hyp$, because then $\hyp$ can be deformed in space-time to a
hypersurface which satisfies \eq{Kres}, while retaining the same
conformal boundary $S$; compare Theorem~\ref{Tposvac} below.

Let us expand on our comments above concerning the Riemannian
definition of mass: consider a CMC initial data set with $\trg K =
-3$ and $\Lambda=0$, corresponding to a hyperboloidal hypersurface
in an asymptotically Minkowskian space-time as constructed
in~\cite{AndChDiss}, so that the $g$-norm of $K_{ij}+g_{ij}$ tends
to zero as one approaches $\scrip$. 
It then follows from the vacuum constraint equations that $R(g)$
approaches $-6$ as one recedes to infinity, and one can enquire
whether the metric satisfies the  conditions needed for the
Riemannian definition of mass for such metrics~\cite{ChHerzlich}.
Now, one of the requirements in~\cite{ChHerzlich} is that the
background derivatives $\znabla g$ of $g$ be in $L^2(M)$. A simple
calculation (see Section~\ref{S3} below) shows that for smoothly
compactifiable $(\hyp,g)$ this will  only be the case if the
extrinsic curvature $\chi$ of the conformally rescaled metric
vanishes at the conformal boundary $\partial \hyp$. It has been
shown in~\cite[Appendix~C.3]{CJK} that the $u$-derivative of
$\chi$ coincides with the Bondi ``news function", and therefore
the Riemannian definition of mass can \emph{not} be used for
families of hypersurfaces in space-times with non-zero flux of
Trautman-Bondi energy, yielding an unacceptable restriction.
Clearly one needs a definition which would allow less stringent
conditions than the ones in~\cite{ChHerzlich,ChNagy}, but this
seems incompatible with the examples in~\cite{ChNagy} which show
sharpness of the conditions assumed. The answer to this apparent
paradox turns out to be the following: in contradistinction with
the asymptotically flat case, in the hyperboloidal one  the
definition of mass does involve the extrinsic curvature tensor $K$
in a non-trivial way. The leading behavior of the latter combines
with the leading behavior of the metric to give a well defined,
convergent, geometric invariant. It is only when $K$ (in the
$\Lambda<0$ case) or $K+g$ (in the $\Lambda=0$ case) vanishes to
sufficiently high order that one recovers the purely Riemannian
definition; however, because the leading order of $K$, or $K+g$,
is coupled to that of $g$ via the constraint equations, one
obtains -- in the purely Riemannian case -- more stringent
conditions on $g-b$ than those arising in the general initial data
context.

We will analyse invariance and finiteness properties of the charge
integrals in any dimension under asymptotic conditions analogous
to those in~\cite{ChNagy,ChHerzlich}. However, the analysis under
boundary conditions appropriate for existence of gravitational
radiation will only be done in space-time dimension
four.

\section{Global charges of initial data sets}
\subsection{The charge integrals} Let $g$ and $b$ be two Riemannian metrics
on an $n$-dimensional manifold $M$, $n\ge 2$, and let $V$ be any
function there. We set\footnote{The reader is warned that the
tensor field $e$ here is \emph{not} a direct Riemannian
counterpart of the one in~\cite{ChNagyATMP}; the latter makes
appeal to the \emph{contravariant} and not the \emph{covariant}
representation of the metric tensor. On the other hand $e_{ij}$
here coincides with that of~\cite{ChHerzlich}.}
\begin{equation}
  \label{eq:3.1}
 e_{ij} :=g_{ij}-b_{ij}\;.
\end{equation}
 We denote
by $\zD$ the Levi-Civita connection of $b$ and by $R_f$ the scalar
curvature of a metric $f$. In~\cite{ChHerzlich} the following
identity has been proved:
\begin{eqnarray}
  \label{eq:3.2} &
  \sqrt{\det g} \;V (R_g-R_b)  =  \partial_i \left(\ourU^i(V)\right) + \sqrt{\det g}
\;(  s + Q)\;, &
\end{eqnarray}
where
\begin{eqnarray} \label{eq:3.3} & {}\ourU^i (V):=  2\sqrt{\det
g}\;\left(Vg^{i[k} g^{j]l} \zD_j g_{kl}
+D^{[i}V 
g^{j]k} e_{jk}\right) \;,&
\\
\label{eq:3.4} & s := (-V\Ric(b)_{ij} +\zD_i \zD_j V -\Delta_b V
b_{ij}) g^{ik} g^{j\ell}
  e_{k\ell}
\;,&
\\
\label{eq:3.5} & Q:= V(g^{ij} - b^{ij} + g^{ik}g^{j\ell}
e_{k\ell})\Ric(b)_{ij} +Q'\;.
\end{eqnarray}
Brackets over a symbol denote anti-symmetrisation, with an
appropriate numerical factor ($1/2$ in the case of two
indices).\footnote{In general relativity a normalising factor
$1/16\pi$, arising from physical considerations, is usually thrown
in into the definition of $\ourU^i$.  From a geometric point view
this seems purposeful when the boundary at infinity is a round two
dimensional sphere; however, for other topologies and dimensions,
this choice of factor  does not seem very useful, and for this
reason we do not include it in $\ourU^i$.} The symbol $\Delta_f$
denotes the Laplace operator of a metric $f$. The result is valid
in any dimension $n\ge 2$. Here $ Q'$ denotes an expression which
is bilinear in $e_{ij}$ and $\zD_k e_{ij}$, linear in $V$, $dV$
and Hess$V$, with coefficients which are uniformly bounded in a
$b$-ON frame, as long as  $g$ is uniformly equivalent to $b$. The
idea behind the calculation leading to \eq{eq:3.2} is to collect
all terms in $R_g$ that contain second derivatives of the metric
in $\partial_i \ourU^i$; in what remains one collects in $s$ the
terms which are linear in $e_{ij}$, while the remaining terms are
collected in $Q$; one should note that the first term at the
right-hand-side of \eq{eq:3.5} does indeed not contain any terms
linear in $e_{ij}$ when Taylor expanded at $g_{ij}=b_{ij}$.

We wish to present a  generalisation of  this formula ---
\Eq{eq:3.2n} below --- which takes into account the physical
extrinsic curvature tensor $K$ and its background equivalent
$\zK$; this requires introducing some notation. For any scalar
field $V$ and vector field $Y$ we define
\begin{eqnarray}
\label{azero} \zA_{kl}\equiv A_{kl}(b,\zK)&:=&\lie_Y b_{kl} - 2V \zK_{kl}\;, \\
  A\equiv A(g,K)&:=& V\underbrace{\left( R_g - K^{kl}K_{kl}
  +(\trg  K)^2\right)}_{=\rho(g,K)}-2Y^k \underbrace{D_l \left(K^l{_k}-\delta^l{_k}\trg  K
  \right)}_{=-J_k(g,K)/2}
  \nonumber \\ &=&
  -\frac{2{\cal G}^0{_\mu} X^\mu}{\sqrt{\det g}}=
 2G_{\mu\nu}n^\mu X^\nu\;.\label{wiaz}
\end{eqnarray}
The symbol $\mycal L$ denotes a Lie derivative.
If we were in a space-time context, then ${\cal G}^\lambda{_\mu}$ 
would be the Einstein tensor density, $G_{\mu\nu}$ would be the
Einstein tensor, while $n^\mu$ would be the future directed normal
to the initial data hypersurface. Finally, an associated
space-time vector field $X$ would then be defined as
\begin{equation}\label{poleX}
 X=Vn^\mu\partial_\mu + Y^k \partial_k = \frac{V}N \partial_0
+(Y^k -\frac{V}N N^k)\partial_k\;,
\end{equation}
where $N$ and $N^k$ are the lapse and shift functions. However, as
far as possible we will forget about any space-time structures.
{It should be pointed out that our $\rho$ here can be interpreted
as the energy-density of the matter fields when the cosmological
constant $\Lambda$ vanishes; it is, however, shifted by a constant
in the general case.}

We set
 \[P^{kl}:=g^{kl}\trg  K- K^{kl}\;, \qquad \trg  K:=g^{kl}K_{kl}\;,\]
with  a similar definition relating the background quantities
$\mathring K$ and $\zP$; indices on $K$ and $P$ are always moved
with $g$ while those on $\zK$ and $\zP$ are always moved with $b$.

We shall say that $(V,Y)$ satisfy the (background) {\em vacuum
Killing Initial Data} (KID) equations if
\bel{a2}A_{ij}(b,\zK) 
=0=S^{kl}(b,\zK) \; ,\ee where
\begin{eqnarray}\label{Skid} 
\zS^{kl}\equiv S^{kl}(b,\zK)& := & \nonumber 
V \left( 2\zP^{ml}\zP_m{}^{k}-\frac3{n-1}\trb \zP\zP^{kl} +
  \Ric(b)^{kl} - R_b b^{kl} \right)   \\ & &
   \mbox{}- \lie_Y \zP^{kl} +\Delta_b V b^{kl} - \zD^k \zD^l V
\;.
\end{eqnarray}
Vacuum initial data with this property lead to space-times with
Killing vectors, see~\cite{Moncrief76}
(compare~\cite{ChBeigKIDs}). We will, however, not assume at this
stage that we are dealing with vacuum initial data sets, and we
will do the calculations in a general case.

 In Appendix~\ref{AVpot} we
derive the following counterpart of \Eq{eq:3.2}:
\begin{eqnarray}
 \partial_i \left(\ourU^i(V)+ \ourV^i(Y)\right)& =&\sqrt{\det g}
\;\left[ V\left(\rho(g,K)-\rho(b,\zK)\right)+ s' +
Q''\right]\nonumber
\\ &&\label{eq:3.2n}
 + Y^k\sqrt{\det g}\,
\left(J_{k}(g,K)-J_k(b,\zK)\right)\;,\nonumber \\
&&
\end{eqnarray}
where
  \bel{matVn} {\mathbb V}^l(Y):=  2\sqrt{\det g}\left[ (P^l{_k}
-\zP^l{_k})Y^k -\frac12
  Y^l\zP^{mn}e_{mn} +\frac12Y^k \zP^l{_k}b^{mn}e_{mn}  \right] \;.
  \ee
   Further, $Q''$ contains
terms which are quadratic in the deviation of $g$ from $b$ and its
derivatives, and in the deviations of $K$ from $\mathring K$,
while $s'$, obtained by collecting all terms linear or linearised
in $e_{ij}$, except for those involving $\rho$ and $J$, reads
$$s'= (\zS^{kl}+\zB^{kl}) e_{kl}+(P^{kl}-\zP^{kl})\zA_{kl}\;,$$
\begin{equation}\label{linw}
\zB^{kl}:= \frac12
  \left[ b^{kl}\zP^{mn}\zA_{mn}
  -b^{mn}\zA_{mn}\zP^{kl}\right]
\;.
\end{equation}

\section{Initial data sets with rapid decay}\label{sec:rd}

\subsection{The reference metrics} \label{S2}
Consider a manifold $M$ which contains a region $\Mext\subset M$
together with a diffeomorphism  \be\label{Nman}\Phi^{-1}:\Mext \to
[R,\infty)\times N\;,\ee where $N$ is a compact boundaryless
manifold. Suppose that on $[R,\infty)\times N$ we are given a
Riemannian metric $b_0$ of a product form \be \label{cm1.0}
b_0:=\frac{dr^2}{r^2+k}+r^2\zh\;, \ee as well as a symmetric
tensor field $K_0$; conditions on $K_0$ will be imposed later on.
We assume that $\zh$ is a Riemannian metric on $N$ with constant
scalar curvature $R_{\zh}$ equal to
 \be\label{cm1.1}
R_{\zh}=
\begin{cases}
(n-1)(n-2)k\;, \quad k\in\{0,\pm 1\} \;, & \mbox{if }\
n>2, \cr 0\;, \quad k=1 \;, & \mbox{if }\ n=2; 
\end{cases}
\ee here $r$ is a coordinate running along the $[R,\infty)$ factor
of $[R,\infty)\times N$. Here the dimension of $N$ is $(n-1)$; we
will later on specialise to the case $n+1=4$ but we allow a
general $n$ in this section. There is some freedom in the choice
of $k$ when $n=2$, associated with the range of the angular
variable $\varphi$ on $N=S^1$, and we make the choice $k=1$ which
corresponds to the usual form of the two-dimensional hyperbolic
space. When $(N,\zh)$  is the unit round $(n-1)$--dimensional
sphere $(\cerc^{n-1},\can)$, then $b_0$ is the hyperbolic metric.

 Pulling-back $b_0$ using
$\Phi^{-1}$ we define on $\Mext$ a reference metric $b$,
\be \label{cm1.} \Phi^*b= b_0\;. \ee
Equations~\eq{cm1.0}--\eq{cm1.1} imply that the scalar curvature
$R_b$ of the metric $b$ is constant:
 $$ R_b=R_{b_0}=
n(n-1)k\;.$$ Moreover, the metric $b$ will be Einstein if and only
if $\zh$ is. We emphasise that for all our purposes we only need
$b$ on $\Mext$, and we continue $b$ in an arbitrary way to
$M\setminus\Mext$ whenever required.

Anticipating, the ``charge integrals"  will be defined as the
integrals of $\ourU + \ourV$ over ``the boundary at infinity",
{\em cf.}\/ Proposition~\ref{P1} below. The convergence of the
integrals there requires appropriate boundary conditions, which
are defined using the following $b$-orthonormal frame
$\{f_i\}_{i=1,n}$ on $\Mext$: \be\label{m1} \Phi^{-1}_*f_i =
r\epsilon_i\;, \quad i=1,\ldots,n-1\;,\quad\Phi^{-1}_*f_n =
\sqrt{r^2+k} \;\partial_r\;, \ee where the $\epsilon_i$'s form an
orthonormal frame for the metric $\zh$. We moreover set \be
\label{m2} g_{ij}:=g(f_i,f_j)\;,\quad K_{ij}:= K(f_i,f_j) \;,\ee
\emph{etc.},\/ and throughout this section only tetrad components
will be used.

\subsection{The charges}
 We start by
introducing a class of boundary conditions for which convergence
and invariance proofs are particularly simple. We emphasise that
the asymptotic conditions of Definition~\ref{adc} are too
restrictive for general hypersurfaces meeting $\scri$ in anti-de
Sitter space-time, or -- perhaps more annoyingly -- for general
radiating asymptotically flat metrics. We will return to that last
case in Section \ref{STB}; this requires considerably more work.

\begin{definition}[{Strong asymptotic decay conditions}] \label{adc}
We shall say that the initial data $(g,K)$
 are \emph{strongly asymptotically hyperboloidal}
if:
\begin{eqnarray}
\lefteqn{\int_{\Mext} \left( \sum_{i,j}\left(
|g_{ij}-\delta_{ij}|^2 + |K_{ij}-\zK_{ij}|^2\right) + \sum_{i,
j,k} |f_k(g_{ij})|^2 \right.}\nonumber  && \\ && \qquad +
\sum_{i,j} \left(|\zS^{ij}+\zB^{ij}|^2 + |\zA_{ij}|^2\right)+\sum_k| J_{k}(g,K)-
J_k(b,\zK)|\nonumber \\
&&\;\left.{\phantom {\sum_k}}+
|\rho(g,K)-\rho(b,\zK)|\right)r\circ\Phi\;d\mu_g<\infty\;,
\label{m3}\end{eqnarray}
\begin{equation}
\label{m0} \exists \ C > 0 \ \textrm{ such that }\ C^{-1}b(X,X)\le
g(X,X)\le Cb(X,X)\;. \end{equation}
\end{definition}

\medskip

Of course, for vacuum metrics $g$ and $b$ (with or without
cosmological constant) and for background KIDs $(V,Y)$ (which will
be mostly of interest to us) all the quantities appearing in the
second and third lines of \eq{m3} vanish.

For hyperboloids in Minkowski space-time, or for static
hypersurfaces in anti de Sitter space-time,
  the $V$'s and $Y$'s associated to the translational
Killing vectors satisfy
\begin{equation}
\label{Vcondition} V=O(r)\;, \quad \sqrt{b^\#(dV,dV)}=O(r)\;,
\qquad |Y|_b=O(r)\;,
\end{equation}
where $b^\#$ is the metric on $T^*M$ associated to $b$, and this
behavior will be assumed in what follows.

 Let $\cNbz$ denote\footnote{We denote by $\zK$ the background extrinsic curvature
 on the physical initial data manifold $M$, and by $\zK_0$ its equivalent in the model manifold $[R,\infty)\times N$,
 $\zK_0:=\Phi^*\zK$.} the
\emph{space of background KIDs}:
\bel{equ:NAME} \cNbz:=\{(V_0,Y_0) \ | \ A_{ij}(b_0,\zK_0) = 0 =
S^{kl}(b_0,\zK_0)\}\;,\ee compare \Eqsone{azero} and \eq{Skid},
where it is understood that $V_0$ and $Y_0$ have to be used
instead of $V$ and $Y$ there. The geometric character of
\eq{azero} and \eq{Skid} shows that if $(V_0,Y_0)$ is a background
KID for $(b_0,\zK_0)$, then $$(V:=V_0\circ \Phi^{-1}, Y:= \Phi_*
Y_0)$$ will be a background KID for $(b=(\Phi^{-1})^*b_0,
\zK=(\Phi^{-1})^*\zK_0)$. The introduction of the $(V_0,Y_0)$'s
provides a natural identification of KIDs for different
backgrounds $((\Phi^{-1}_1)^*b_0, (\Phi^{-1}_1)^*\zK_0)$ and
$((\Phi^{-1}_2)^*b_0, (\Phi^{-1}_2)^*\zK_0)$ . We have

\begin{pro}
\label{P1} Let the reference metric $b$ on $\Mext$ be of the form
\eq{cm1.}, suppose that $V$ and $Y$ satisfy \eq{Vcondition}, and
assume that $\Phi$ is such that Equations \eq{m3}-\eq{m0} hold.
Then for all $(V_0,Y_0)\in\cNbz$ the limits \be \label{mi}
H_\Phi(V_0,Y_0):=\lim_{R\to\infty} \int_{r\circ \Phi^{-1}=R}
\left(\ourU^i(V_0\circ
\Phi^{-1})+\ourV^i(\Phi_* Y_0 
)\right) dS_i
\ee exist, and are finite.
\end{pro}

The integrals~\eq{mi} will  be referred to as \emph{the Riemannian
charge integrals}, or simply \emph{charge integrals}.

\medskip

\proof We work in coordinates on $\Mext$ such that $\Phi$ is the
identity. For any $R_1,R_2$ we have
\be \int_{r=R_2} (\ourU^i+\ourV^i) dS_i = \int_{r=R_1}
(\ourU^i+\ourV^i) dS_i + \int_{[R_1,R_2]\times N}
\partial_i (\ourU^i+\ourV^i) \,d^n x\;, \ee and the result follows from
\eq{eq:3.2n}-\eq{linw}, together with \eq{m3}-\eq{m0} and the
Cauchy-Schwarz inequality, by passing to the limit $R_2\to
\infty$. \myqed

\medskip

In order to continue we need some more restrictions on the
extrinsic curvature tensor $\zK$. In the physical applications we
have in mind in this section the tensor field  $\zK$ will be pure
trace, which is certainly compatible with the following
hypothesis:
\bel{m5.0} |\zK^{i}{}_{j}-\frac{\trz \zK}n
\delta^i_j|_{b_0}=o(r^{-n/2})\;.\ee Under \eq{m5.0} and \eq{m5}
one easily finds from \eq{matVn} that \bel{m5.0a}
\lim_{R\to\infty} \int_{r\circ \Phi^{-1}=R} \ourV^i(Y)
dS_i=2\lim_{R\to\infty} \int_{r\circ \Phi^{-1}=R}\sqrt{\det
b}\left[ (P^l{_k} -\zP^l{_k})Y^k \right] dS_i\;,\ee which gives a
slightly simpler expression for the contribution of $P$ to
$H_\Phi$.

 Under the
conditions of Proposition~\ref{P1}, the integrals~\eq{mi} define a
linear map from $\cNbz$ to $\R$. Now, each map $\Phi$ used in
\eq{cm1.} defines in general a different background metric $b$ on
$\Mext$, so that the maps $H_\Phi$ are potentially
dependent\footnote{Note that the space of KIDs is fixed, as
$\cNbz$ is tied to $(b_0,\zK_0)$ which are fixed once and for
all.} upon $\Phi$. (It should be clear that, given a fixed $\zh$,
\eq{mi} does not depend upon the choice of the frame $\epsilon_i$
in \eq{m1}.) It turns out that this dependence can be controlled:
\newcommand{\AY}{A_*Y_0}
\begin{theo}\label{Tinv} Under \eq{m5.0},
consider two maps $\Phi_a$, $a=1,2$, satisfying \eq{m3} together
with
 \begin{eqnarray}\nonumber
\lefteqn{\sum_{i,j}
\left(|g_{ij}-\delta_{ij}|+|P^i{}_{j}-\zP^i{}_{j}|\right)
+ \sum_{i, j,k} |f_k(g_{ij})| =}&&\\
&&\mbox{}= \begin{cases}
o(r^{-n/2})\;, & \mbox{if }\ n>2, \cr O(r^{-1-\epsilon})\;, &
\mbox{if }\ n=2,\ \mbox{for some $\epsilon>0$}.
\end{cases} \label{m5}\end{eqnarray} Then there exists an isometry $A$ of $b_0$, defined
perhaps only for $r$ large enough, such that \bel{Htransf}
H_{\Phi_2}(V_0,Y_0)= H_{\Phi_1}\left(V_0\circ A^{-1}, \AY
\right)\;. \ee
\end{theo}

\begin{remk}\label{RTinv}The examples in~\cite{ChNagyATMP} show
that the decay rate \eq{m5} is sharp when $\zP_{ij}=0$, or when
$Y^i=0$, compare~\cite{ChHerzlich}.
\end{remk}
 \proof When $\zK=0$ the result is proved, using a space-time formalism, at
  the beginning of Section~4 in~\cite{ChNagyATMP}.
When $Y=0$ this
  is Theorem~2.3
of~\cite{ChHerzlich}. It turns out that under~\eq{m5.0} the
calculation reduces to the one in that last theorem, and that
under the current conditions the integrals of $\ourU$ and $\ourV$
are separately covariant, which can be seen as follows: On $\Mext$
we have three pairs of fields:
$$(g,K)\;,\quad  \Big((\Phi^{-1}_1)^*b_0,
(\Phi^{-1}_1)^*\zK_0\Big)\;,\ \mbox{ and } \
\Big((\Phi^{-1}_2)^*b_0, (\Phi^{-1}_2)^*\zK_0\Big)\;.$$ Pulling
back everything by $\Phi_2$ to $[R,\infty)\times N$ we obtain
there
$$\Big((\Phi_2)^*g,(\Phi_2)^*K\Big)\;,\quad \Big((\Phi^{-1}_1\circ \Phi_2)^*b_0, (\Phi^{-1}_1\circ
\Phi_2)^*\zK_0\Big) \;,\  \mbox{ and } \ (b_0, \zK_0)\;.$$ Now,
$(\Phi_2)^*g$ is simply ``the metric $g$ as expressed in the
coordinate system $\Phi_2$", similarly for $(\Phi_2)^*K$, and
following the usual physicist's convention we will instead write
$$(g,K)\;,\quad  (b_1,\zK_1):=\Big((\Phi^{-1}_1\circ \Phi_2)^*b_0,
(\Phi^{-1}_1\circ \Phi_2)^*\zK_0\Big)\;,\  \mbox{ and } \
(b_2,\zK_2)=(b_0, \zK_0)\;,$$ which should be understood in the
sense just explained.

 As discussed in more detail in~\cite[Theorem~2.3]{ChHerzlich}, there exists an isometry $A$ of
the background metric $b_0$, defined perhaps only for $r$ large
enough, such that $\Phi^{-1}_1\circ \Phi_2$ is a composition of
$A$ with a map which approaches the identity as one approaches the
conformal boundary, see \eq{eqnm5a}-\eq{eqnm5} below. It can be
checked that the calculation of the proof
of~\cite[Theorem~2.3]{ChHerzlich} remains valid, and yields
\bel{Htransf2} H_{\Phi_2}(V_0,0)= H_{\Phi_1}(V_0\circ A^{-1},0)\;. \ee
(In~\cite[Theorem~2.3]{ChHerzlich} \Eq{Skid} with $Y=0$ has been
used. However, under the hypothesis \eq{m5.0} the supplementary
terms involving $Y$ in \eq{m5.0} cancel out in that calculation.)
It follows directly from the definition of $H_{\Phi}$ that
\be H_{\Phi_1\circ A}(0,Y_0)= H_{\Phi_1}\left(0, \AY
\right)\;.\ee
Since
$$H_{\Phi}(V_0,Y_0)=H_\Phi(V_0,0)+H_\Phi(0,Y_0)\;,$$
we need  to show that
\bel{Htransf3} H_{\Phi_2}(0,Y_0)= H_{\Phi_1}\left(0, \AY  \right)\;. \ee
 In order
to establish \eq{Htransf3} it remains to show that for all $Y_0$
we have
\bel{Htransf2.1} H_{\Phi_1\circ A} (0,Y_0)= H_{\Phi_2}(0,Y_0)\;.
\ee Now, Corollary~3.5 of \cite{ChNagyATMP} shows that the
pull-back of the metrics by $\Phi_1\circ A$ has the same decay
properties as that by $\Phi_1$, so that
--- replacing $\Phi_1$ by $\Phi_1\circ A$ --- to prove
\eq{Htransf2.1} it remains to consider two maps $\Phi_1^{-1}
=(r_1,v^A_1)$ and $\Phi_2^{-1}=(r_2,v^A_2)$ (where $v^A$ denote
abstract local coordinates on $N$) satisfying
\begin{eqnarray}
r_2 & = r_1 + o(r_1^{1-\frac{n}{2}}) ,
\label{eqnm5a}\\
v^A_2 & = v^A_1 + o(r_1^{-(1+\frac{n}{2})}) ,
\label{eqnm5}
\end{eqnarray}
together with similar derivative bounds. In that case one has, in
tetrad components, by elementary calculations,
\bel{Ptransf}P{}^i{}_j - \zP_1{}^i{}_j = P{}^i{}_j - \zP_2{}^i{}_j
+o(r^{-n})\;, \ee leading immediately to \eq{Htransf2.1}. We point
out that it is essential that $P^i{}_j$ appears in \eq{Ptransf}
with one index up and one index down. For example, the difference
$$P_{ij} - \zP_{ij} =o(r^{-n/2})\;,
$$
transforms as
$$P_{ij} - \zP_1{}_{ij}= P_{ij} - \zP_2{}_{ij} -
\tr_{b_1}\zP_1 \mathcal{L}_{\zeta}b_1+o(r^{-n})\;,
$$
where \[ \zeta = (r_2 - r_1)\frac{\partial}{\partial r_1} + \sum_A
(v^A_2 - v^A_1) \frac{\partial}{\partial v^A_1} \ . \] \myqed
\medskip

%

\section{Problems with the extrinsic curvature of the conformal boundary}
\label{S3}

Consider a vacuum space-time $(\mcM,\fourg)$, with cosmological
constant $\Lambda=0$, which possesses a smooth conformal
completion $(\bmcM,\tfg)$ with conformal boundary $\scrip$.
Consider a hypersurface $\hyp$ such that its completion $\thyp$ in
$\bmcM$ is a smooth spacelike hypersurface intersecting $\scrip$
transversally, with $\thyp\cap\scrip$ being smooth two-dimensional
sphere;  no other completeness conditions upon  $\mcM$, $\strut\bmcM$,
or upon $\scrip$ are imposed. In a neighborhood of
$\thyp\cap\scrip$ one can introduce Bondi
coordinates~\cite{Tamburino:Winicour}, in terms of which $\fourg$
takes the form
\be\label{gB0} \fourg= -xV{\rm
e}^{2\beta}\,d u^2 + 2 {\rm e}^{2\beta} x^{-2}\,d u \,d x
   + x^{-2} h_{AB} \left(\,d x^A - U^A\,d u \right)
   \left(\,d x^B - U^B\,d u \right)\;. \ee
   (The usual radial Bondi coordinate $r$ equals $1/x$.)
One has \bel{gb1}  h_{AB}=\hst _{AB}+x\chi _{AB}+O(x^{2})\;,\ee
where $\hst$ is the round unit metric on $S^2$, and the whole
information about gravitational radiation is encoded in the tensor
field $\chi_{AB}$. It has been shown in~\cite[Appendix~C.3]{CJK}
that the trace-free part of the extrinsic curvature of
$\thyp\cap\scri$
 within $\hyp$ is proportional to $\chi$. In coordinate systems
on $\hyp$ of the kind used in \eq{cm1.0} this leads to a $1/r$
decay of the tensor field $e$ of \eq{eq:3.1}, so that the decay
condition \eq{m3} is not satisfied. In fact, \eq{eq:3.1} ``doubly
fails" as the $K-\zK$ contribution also falls off too slowly for
convergence of the integral. Thus the decay conditions of
Definition~\ref{adc} are not suitable for the problem at hand.

Similarly, let $\hyp$ be a space-like hypersurface in
 a vacuum space-time $(\mcM,\fourg)$ with strictly negative
cosmological constant $\Lambda$, with a smooth conformal
completion $(\tM,\tfg)$ and conformal boundary $\scri$, as
considered {\em e.g.\/} in~\cite[Section~5]{ChruscielSimon}. Then
a generic smooth deformation of $\hyp$  at fixed conformal
boundary $\thyp\cap\scri$ will lead to induced initial data which
will not satisfy \eq{m3}.

 As already
mentioned, in some of the calculations we will not consider the
most general hypersurfaces compatible with the set-ups just
described, because the calculations required seem to be too
formidable to be performed by hand. We will instead impose the
following restriction:
\bel{fr1}
d(\tr_gK) \ \mbox{ vanishes on the conformal boundary. }\ee In
other words, $\tr_gK$ is constant on $\thyp\cap\scri$, with the
transverse derivatives of $\tr_gK$ vanishing there as well.

\Eq{fr1} is certainly a restrictive assumption. We note, however,
the following: \begin{enumerate} \item It holds for all initial
data constructed by the conformal method, both if
$\Lambda=0$~\cite{AndChDiss} and if $\Lambda<0$~\cite{Kannar:adS},
for then $d(\trg K)$ is zero throughout $\hyp$.
\item One immediately sees from the equations
in~\cite[Appendix~C.3]{CJK} that in the case $\Lambda=0$ \Eq{fr1}
can be achieved by deforming $\hyp$ in $\mcM$, while keeping
$\thyp\cap \scrip$ fixed, whenever an associated space-time
exists. It follows that for the proof of positivity of $m$ in
space-times with a conformal completion it suffices to consider
hypersurfaces satisfying \eq{fr1}.
\end{enumerate}

\section{Convergence, uniqueness, and positivity of the \\Trautman-Bondi mass}
\label{STB}

{}From now on we assume that the space dimension is three, and
that
$$\Lambda = 0\;.$$ An extension  to higher dimension would require
studying Bondi expansions in $n+1>4$, which appears to be quite a
tedious undertaking. On the other hand the adaptation of our
results here to the case $\Lambda<0$ should be straightforward,
but we have not attempted such a calculation.


 The metric $g$ of a
Riemannian manifold $(M,g)$ will be said to be {\em $C^k$
compactifiable} if there exists a compact Riemannian manifold with
boundary $(\bM\approx M\cup \piM\cup
\partial M ,\tg)$, where
$\partial\bM=\partial M\cup\piM$ is the metric boundary of
$(\bM,\tg)$, with  $\partial M$ --- the metric boundary of
$(M,g)$, together with a diffeomorphism
$$\psi: \Int \bM\to M$$ such that \be \label{ccond} \psi^*g=
x^{-2} \tg\;,\ee where $x$ is a defining function for $\piM$ ({\em
i.e.,\/} $x\ge 0$, $\{x=0\}=\piM$, and $dx$ is nowhere vanishing
on $\piM$), with $\tg$ --- a metric which is $C^k$
up--to--boundary on $\bM$. The triple $(\bM,\tg,x)$ will then be
called a \emph{$C^k$ conformal completion} of $(M,g)$. Clearly the
definition allows $M$ to have a usual compact boundary. $(M,g)$
will be said to have a \emph{conformally compactifiable end
$\Mext$} if $M$ contains an open submanifold $\Mext$ (of the same
dimension that $M$) such that $(\Mext,g|_{\Mext})$ is conformally
compactifiable, with a connected conformal boundary
$\partial_\infty \Mext$.

In the remainder of this work we shall assume for simplicity that
the conformally rescaled metric $\tg$ is {\em polyhomogeneous} and
$C^1$   near the conformal boundary; this means that $\tg$ is
$C^1$ up-to-boundary and has an asymptotic expansion with smooth
expansion coefficients to any desired order in terms of powers of
$x$ and of $\ln x$. (In particular, smoothly compactifiable
metrics belong to the polyhomogeneous class; the reader unfamiliar
with polyhomogeneous expansions might wish to assume smoothness
throughout.) It should be clear that the conditions here can be
adapted to metrics which are polyhomogeneous plus a weighted
H\"older or Sobolev lower order term decaying sufficiently fast.
In fact, a very conservative estimate, obtained by inspection of
the calculations below, shows that relative $\Ol(x^4)$ error terms
introduced in the metric because of matter fields or because of
sub-leading non-polyhomogeneous behavior do not affect the
validity of the calculations below, provided the derivatives of
those error term behave under differentiation in the obvious way
(an $x$ derivative lowers the powers of $x$ by one, other
derivatives preserve the powers). (We use the symbol $f=\Ol(x^p)$
to denote the fact that there exists $N\in \N$ and a constant $C$
such that $|f|\le Cx^p(1+|\ln x|^N)\;.$)

 An initial data set $(M,g,K)$ will be said
to be $C^k(\bM)\times C^\ell(\bM)$ {\em conformally
compactifiable} if $(M,g)$ is  $C^k(\bM)$ conformally
compactifiable  and if $K$ is of the form
\bel{decompL} K^{ij}=x^3 L^{ij} + \frac {\trg K}{3}
g^{ij}\;,\ee with the trace-free tensor $L^{ij}$  in
$C^\ell(\bM)$, and with $\trg K$ in $C^\ell(\bM)$, strictly
bounded away from zero on $\bM $.  We note that \eq{m5} would have
required $|L|_{\tg}=o(x^{1/2})$, while we allow $|L|_{\tg}=O(1)$.
The slower decay rate is necessary in general for compatibility
with the constraint equations if the trace-free part of the
extrinsic curvature tensor of the conformal boundary does not
vanish (equivalently, if the tensor field $\chi_{AB}$ in \eq{gb1}
does not vanish); this follows from the calculations in
Appendix~\ref{APLpos1}.

 \subsection{The Trautman-Bondi four-momentum of asymptotically \\
hyperboloidal initial data sets -- the four-dimensional \\
definition}\label{Sdef4}

The definition \eq{mi} of global charges requires  a background
metric $b$, a background extrinsic curvature tensor $\zK$, and a
map $\Phi$. For initial data which are vacuum near $\scrip$ all
these objects will now be defined using Bondi coordinates, as
follows: Let $(\hyp,g,K)$ be a hyperboloidal initial data set, by
\cite{ChLengardprep,Lengard} the associated vacuum space-time
$(\mcM,\fourg)$ has a conformal completion $\scrip$, with perhaps
a rather low degree of differentiability. One expects that $\spt$
will indeed be polyhomogeneous, but such a result has not been
established so far. However, the analysis of~\cite{ChMS} shows
that one can formally determine all the expansion coefficients of
a polyhomogeneous space-time metric {\em on $\hyp$}, as well as
all their time-derivatives on $\hyp$. This is sufficient to carry
out all the calculations here {\em as if the resulting completion
were polyhomogeneous}. In all our calculations from now on we
shall therefore assume that $\spt$ has a polyhomogeneous conformal
completion,  this assumption being understood in the sense just
explained.

 In $\spt$ we can
always~\cite{ChMS} introduce a Bondi coordinate system $(u,x,x^A)$
such that $\hyp$ is given by an equation \bel{a1n} u=\alpha(x,x^A)
\;,  \ \mbox{ with } \ \alpha(0,x^A)=0\;, \ \alpha_{,x}(0,x^A)
> 0\;,\ee where $\alpha$ is polyhomogeneous. There is \emph{exactly} a
six-parameter family of such coordinate systems, parameterised by
the Lorentz group (the supertranslation freedom is gotten rid of
by requiring that $\alpha$ vanishes on $\thyp\cap \scri$). We use
the Bondi coordinates to define the background $\fourb$:
\bel{fourb} \fourb := -\rd u^2 +2x^{-2} \rd u \rd x +
x^{-2}\hst_{AB}\rd x^A \rd x^B \;.\ee This Lorentzian background
metric $\fourb$ is independent of the choice of Bondi coordinates
as above. One then defines the Trautman-Bondi four-momentum
$p_\mu$ of the asymptotically hyperboloidal initial data set we
started with as the Trautman-Bondi four-momentum of the cut $u=0$
of the resulting $\scrip$; recall that the latter is defined as
follows: Let $X$ be a translational Killing vector of $\fourb$, it
is shown \emph{e.g.} in \cite[Section~6.1, 6.2 and 6.10]{CJK} that
the integrals
$$H(\hyp, X,\fourg,\fourb):=\lim _{\epsilon\to 0} \int_{\{x=\epsilon\}\cap \hyp}
\ourW^{\nu\lambda}(X,\fourg,\fourb)dS_{\nu\lambda} $$ converge.
Here $ \ourW^{\nu\lambda}(X,\fourg,\fourb)$ is given by
\eq{Fsup2new}. Choosing an ON basis $X_\mu$ for the $X$'s one then
sets
$$p_\mu(\hyp) := H(\hyp, X_\mu,\fourg,\fourb)\;.$$
(The resulting numbers coincide with the Trautman-Bondi
four-momentum; we emphasise that the whole construction depends
upon the use of Bondi coordinates.)

\subsection{Geometric invariance}\label{Sginv}

 The definition just given involves two arbitrary elements:
the first is the choice of a conformal completion, the second is
that of a Bondi coordinate system. While the latter is easily
taken care of, the first requires attention. Suppose, for example,
that a prescribed region of a space-time $(\mcM,g)$ admits two
completely unrelated conformal completions, as is the case for the
Taub-NUT space-time. In such a case the resulting $p_\mu$'s might
have nothing to do with each other. Alternatively, suppose that
there exist two conformal completions which are homeomorphic but
not diffeomorphic. Because the objects occurring in the definition
above  require derivatives of various tensor fields, one could a
priori again obtain different answers. In fact, the construction
of the approximate Bondi coordinates above requires expansions to
rather high order of the metric at $\scrip$, which is closely
related to high differentiability of the metric at $\scrip$, so
even if we have two diffeomorphic completions such that the
diffeomorphism is not smooth enough, we might still end up with
unrelated values of $p_\mu$.

It turns out that none of the above can happen. The key element of
the proof is the following result, which is essentially
Theorem~6.1 of~\cite{ChHerzlich}; the proof there was given for
$C^\infty$ completions, but an identical argument applies under
the hypotheses here:

\begin{theo}\label{Tucc} Let $(M,g)$ be a Riemannian manifold endowed with two $C^k$,
\mbox{$k\ge 1$}  and polyhomogeneous conformal compactifications
$(\bM_1,\tg_1,x_1)$ and \linebreak $(\bM_2,\tg_2,x_2)$ with compactifying
maps $\psi_1$ and $\psi_2$. Then
$$\psi_1^{-1}\circ\psi_2: \Int M_2\to \Int M_1 $$
extends by continuity to a $C^k$ and polyhomogeneous conformal
up-to-boundary diffeomorphism from $(\bM_2,\tg_2)$ to
$(\bM_1,\tg_1)$, in particular $\bM_1$ and $\bM_2$ are
diffeomorphic as manifolds with boundary.
\end{theo}

We are ready now to prove definitional uniqueness of
four-momentum. Some remarks are in order: \begin{enumerate} \item
It should be clear that the proof below generalises to matter
fields near $\scrip$ which admit a well posed conformal Cauchy
problem \emph{\`a la} Friedrich, {\em e.g.}\/ to
Einstein-Yang-Mills fields~\cite{HelmutJDG}.
\item The differentiability conditions below have been chosen to
ensure that the conformal Cauchy problem of
Friedrich~\cite{Friedrich:grg15} is well posed; we have taken a
very conservative estimate for the differentiability thresholds,
and for simplicity we have chosen to present the results in terms
of classical rather than Sobolev differentiability. One expects
that $C^1(\bM)\times C^0(\bM)$ and polyhomogeneous CMC initial
data $(\tg, L) $ will lead to existence of a polyhomogeneous
$\scrip$; such a theorem would immediately imply a corresponding
equivalent of Theorem~\ref{Tmuniq}.
\item It is clear that there exists a purely three-dimensional version of the proof below, but we have not attempted
to find one; the argument given seems to minimise the amount of
new calculations needed. Such a three-dimensional proof would
certainly provide a result under much weaker asymptotic conditions
concerning both the matter fields and the requirements of
differentiability at $\scrip$.
\end{enumerate}

\begin{theo}\label{Tmuniq} Let $(M,g,K)$ be a
 $C^7(\bM)\times C^6(\bM)$ and polyhomogeneous conformally
compactifiable initial data set which is vacuum near the conformal
boundary, and consider two $C^7(\bM)$ and polyhomogeneous
compactifications thereof as in Theorem~\ref{Tucc}, with
associated four-momenta $p_\mu^a$, $a=1,2$. Then there exists a
Lorentz matrix $\Lambda_\mu{}^\nu$ such that
$$p^1_\mu =\Lambda_\mu{}^\nu p^2_\nu\;.$$
\end{theo}

\proof By the results of Friedrich (see~\cite{Friedrich:grg15} and
references therein) the maximal globally hyperbolic development
$(\mcM,\fourg)$ of $(M,g,K)$ admits $C^4$ conformal completions
$(\bmcM_a,\bfourg_a)$, $a=1,2$ with conformal factors $\Omega_a$
and diffeomorphisms $\Psi_a:\mathrm{int}\; \bmcM_a\to \mcM$ such
that
$$\Psi_a^*(\fourg) = \Omega_a^{-2}\; \bfourg_a\;,\quad \Psi_a|_M =
\psi_a\;.$$
The uniqueness-up-to-conformal-diffeomorphism property
of the conformal equations of Friedrich together with
Theorem~\ref{Tucc} show that $\Psi^{-1}_2\circ \Psi_1$ extends by
continuity to a $C^4$-up-to-boundary map from a neighborhood of
$\Psi_a^{-1}(\bM)\subset \bmcM_1$ to $\bmcM_2$. Let $b_a$ be the
Minkowski background metrics constructed near the respective
conformal boundaries $\scrip_a$ as in Section~\ref{Sdef4},  we
have
$$ \left(\Psi_1\circ \Psi_2^{-1}\right)^* b_2=b_1\;, $$
so that $\left(\Psi_1\circ \Psi_2^{-1}\right)^*$ defines a Lorentz
transformation between the  translational Killing vector fields of
$b_1$ and $b_2$, and the result follows \emph{e.g.}\/
from~\cite[Section~6.9]{CJK}. \qed

\subsection{The Trautman-Bondi four-momentum of asymptotically \\ CMC
 hyperboloidal initial data sets -- a
three-dimensional definition} \label{STBah}


Consider a conformally compactifiable initial data set $(M,g,K)$
as defined in Section~\ref{STB}, see \eq{decompL}. We shall say
that
 $(M,g,K)$ is
 {\em asymptotically CMC\/} if $\trg K$ is in $C^1(\bM )$ and if
  \be\label{cm2hb} \mbox{the differential of $\trg K$ vanishes on
$\piMx$}\;.
 \ee
The vacuum scalar constraint equation ($\rho=0$ in \eq{wiaz})
shows that, for $C^1(\bM)\times C^1(\bM)$ (or for $C^1(\bM)\times
C^0(\bM)$ and polyhomogeneous) conformally compactifiable initial
data sets,  \Eq{cm2hb} is equivalent to
\be\label{cm2h} \mbox{the differential of the Ricci scalar $R(g)$
vanishes on $\piMx$}\;.
 \ee

 We wish, now, to show that for asymptotically CMC
initial data sets one can define a mass in terms of limits
\eq{mi}. The construction is closely related to that presented in
Section~\ref{Sdef4}, except that everything will be directly read
off from the initial data: \emph{If} a space-time as in
  Section~\ref{Sdef4}  exists, then we define the Riemannian background metric
$b$ on $\hyp$  as the metric induced by the metric $\fourb$ of
Section~\ref{Sdef4} on the hypersurface $u=\alpha$, and $\zK$ is
defined as the extrinsic curvature tensor, with respect to
$\fourb$, of that hypersurface. The map $\Phi$ needed in \eq{mi}
is defined to be the identity in the Bondi coordinate system
above, and the metric $b_0$ is defined to coincide with $b$ in the
coordinate system above. The four translational Killing vectors
$X_\mu$ of $\fourb$ induce on $\hyp$ four KIDs $(V,Y)_\mu$, and
one can plug those into \eq{mi} to obtain a definition of
four-momentum. \emph{However},\/ the question of existence and/or
of construction of the space-time there is completely circumvented
by the fact that \emph{the asymptotic development of the function
$\alpha$, and  that of $b$, can be read off directly from $g$ and
$K$}, using the equations of~\cite[Appendix~C.3]{CJK}. The method
is then to read-off the restrictions $x|_\hyp$ and $x^A|_\hyp$ of
the space-time Bondi functions $x$ and $x^A$ to $\hyp$ from the
initial data, and henceforth the asymptotic expansions of all
relevant Bondi quantities in the metric, up to error terms
$\Ol(x^4)$ (order $O(x^4)$ in the smoothly compactifiable case);
equivalently, one needs an approximation of the Bondi coordinate
$x$ on $\hyp$ up to error terms $\Ol(x^5)$. The relevant
coefficients can thus be recursively read from the initial data by
solving a finite number of recursive equations. The resulting
approximate Bondi function $x$ induces a foliation of a
neighborhood of the conformal boundary, which will be called the
\emph{approximate Bondi foliation}. The asymptotic expansion of
$\alpha$ provides an identification of $\hyp$ with a hypersurface
$u=\alpha$ in Minkowski space-time with the flat metric
\eq{fourb}. The Riemannian background metric $b$ is defined to be
the metric induced by $\fourb$ on this surface, and $\Phi$ is
defined to be the identity in the approximate Bondi coordinates.
As already indicated, the KIDs are obtained on $\hyp$ from the
translational Killing vector fields of $\fourb$. The charge
integrals \eq{mi} have to be calculated \emph{on the approximate
Bondi spheres} $x=\epsilon$ before passing to the limit
$\epsilon\to0$.

The simplest question one can ask is whether the linearisation of
the integrals \eq{mi} reproduces the linearisation of the Freud
integrals under the procedure above. We show in
Appendix~\ref{Slin} that this is indeed the case. We also show in
that appendix that the linearisation of the Freud integrals
\emph{does not} reproduce the Trautman-Bondi four-momentum in
general, see \Eq{badeq}. On the other hand that linearisation
provides the right expression for $p_\mu$ when the extrinsic
curvature $\chi$ of the conformal boundary vanishes. Note that
under the boundary conditions of Section~\ref{sec:rd} the
extrinsic curvature $\chi$ vanishes, and the values of the charge
integrals coincide with the values of their linearised
counterparts for translations, so that the calculations in
Appendix~\ref{Slin} prove the equality of the $3+1$ charge
integrals and the Freud ones for translational background KIDs
under the conditions of Section~\ref{sec:rd}.

 The main result of this section is the following:

\begin{theo}
\label{Tconv2} Consider an asymptotically CMC initial data set
which is $C^1$ and polyhomogeneously (or smoothly) conformally
compactifiable. Let $\Phi$ be defined as above and let $(V,Y)_\mu$
be the background KIDs associated to space-time translations
$\partial_\mu$. Then the limits \eq{mi}
$H_\Phi\left((V,Y)_\mu\right)$ taken along \emph{approximate Bondi
spheres} $\{x=\epsilon\}\subset \hyp$ exist and are finite.
Further, the numbers \bel{mtbd} p_\mu :=
H_\Phi\left((V,Y)_\mu\right)\ee coincide with  the {\em
Trautman-Bondi four-momentum} of the associated cut in the
Lorentzian space-time, whenever such a space-time exists.
\end{theo}

\proof  We will show that $\ourU^x+\ourV^x$ coincides with
\eq{int2.1} below up to a complete divergence and up to lower
order terms not contributing in the limit, the result follows then
from \eq{etb}-\eq{ptb}. It is convenient to rewrite the last two
terms in \eq{matVn} as
\bel{toto} -\frac12 Y^x\zP^{m}{}_{n}e^m{}_{n} +\frac12Y^k
\zP^x{_k}b^{mn}e_{mn}\;, \ee so that we can use \eq{PP0}-\eq{PPA}
with $M=\chi_{AB}=\beta=N^A=0$ there. We further need the
following expansions (all indices are coordinate ones)
\begin{eqnarray*}
e^k{_l} &:= &b^{km} (g_{ml}-b_{ml})\;, \\
e^x{_x} & = & 2\beta + x^3 \alpha_{,x} M + \Ol(x^4) \;, \\
e^x{_A} & = & \frac14 x^2 \chi_{AC}{^{||C}} +
\frac{\alpha_{,A}}{2\alpha_{,x}}
- x^3 N_A - \frac{1}{32} x^3 (\chi^{CD} \chi_{CD})_{||A} + \Ol(x^4) \;, \\
e^A{_x} & = & \frac12 x^2 \alpha_{,x} \chi^{AC}{_{||C}} +
\alpha^{,A} - 2x^3\alpha_{,x}\left( N^A + \frac{1}{32}(\chi^{CD} \chi_{CD})^{||A}\right)
 \! + \Ol(x^4) \;, \\
e^A{_B} & = & x \chi^A{_B} + \frac 14 x^2 \chi^{CD}\chi_{CD}
\delta^A{_B} + x^3 \xi^A{_B} + \Ol(x^4) \;,
\end{eqnarray*}
with $||$ denoting a covariant derivative with respect to $\hst$.
One then finds
\begin{eqnarray*}
-\frac12 Y^x\zP^{mn}e_{mn} +\frac12Y^k \zP^x{_k}b^{mn}e_{mn} & = &
Y^x \cdot \Ol(x^4) + Y^B \cdot \Ol(x^5) \;, \\
-\frac12 Y^A\zP^{mn}e_{mn} +\frac12Y^k \zP^A{_k}b^{mn}e_{mn} & = &
Y^x \cdot \Ol(x^5) + Y^B \cdot \Ol(x^4) \;. \\
\end{eqnarray*}
This shows that for $Y^i$ which are  $O(1)$ in the $(x,x^A)$
coordinates, as is the case here (see
Appendix~\ref{app:killingi}), the terms above multiplied by
$\sqrt{\det g} = O(x^{-3})$ will give zero contribution in the
limit, so that in \eq{matVn} only the first two terms will
survive. Those are clearly equal to the first two terms in
\eq{int2.1} when a minus sign coming from the change of the
orientation of the boundary is taken into account.

On the other hand  $\ourU^x$ does not coincide with the remaining
terms in \eq{int2.1},  instead with some work one finds
\begin{eqnarray}\nonumber
2(\lambda k - \zlambda \zk)V - \ourU^x &=& V\Big[\sqrt{\det g}
\cdot \gtw^{AB} g^{xm} \zD_A g_{mB} \\\nonumber && + 2
\Big(\sqrt{\frac{b^{xx}}{g^{xx}} } -1\Big)\zlambda \zk +
2\sqrt{\det g} (\gtw^{AB} - \btw^{AB}) {\zGamma}^{x}_{AB}
\Big]\\\nonumber && - 2\sqrt{\det g} D^{[x}Vg^{j]k}e_{jk}
\\ &=& V \sqrt{\hst } \left(
\frac{x}4\chi^{AB}{_{||AB}}+O(x^2)\right)\;, \label{masssphere}
\end{eqnarray}
where $\gtw^{AB} g_{BC} =
\delta^A{}_C$ and similarly for $\btw^{AB}$. However, integration
over $S^2$ yields equality in the limit; here one has to use the
fact that $V$ is a linear combination of $\ell=0$ and $1$
spherical harmonics in the relevant order in $x$ (see
Appendix~\ref{app:killingi}), so that the trace-free part of
$V{_{||AB}}$ is $ O(x)$.
 \qed

\subsection{Positivity}\label{Sspos}

Our next main result is the proof of positivity of the
Trautman-Bondi mass:

\begin{theo}
\label{Tpos} Suppose that $(M,g)$ is geodesically complete without
boundary. Assume that $(M,g,K)$ contains an end which is
$C^4\times C^3$, or $C^1$ and polyhomogeneously, compactifiable
and asymptotically CMC. If
\bel{dec}\sqrt{g_{ij}J^iJ^j} \le \rho\in L^1(M)\;,\ee then $p_\mu$ is {\em timelike
future directed or vanishes}, in the following sense:
\bel{pmineq}p_0 \ge \sqrt{\sum p_i^2}\;.\ee Further, equality
holds if and only if there exists a $\nabla$-covariantly constant
spinor field on $M$.
\end{theo}

\medskip

\begin{Remark}
\label{RTpos2} {\rm We emphasise that no assumptions about the
geometry or the behavior of the matter fields except geodesic
completeness of $(M,g)$ are made on $M\setminus \Mext$.}
\end{Remark}

\begin{Remark}
\label{RTpos1} {\rm In vacuum one expects that equality in
\eq{pmineq} is possible only if the future maximal globally
hyperbolic development of $(M,g,K)$ is isometrically diffeomorphic
to a subset of the Minkowski space-time; compare~\cite{ChBeig1}
for the corresponding statement for initial data which are
asymptotically flat in space-like directions. When $K$ is pure
trace one can use a result of Baum~\cite{Baum} to conclude that
$(M,g)$ is the three-dimensional hyperbolic space, which implies
the rigidity result. A corresponding result with a general $K$ is
still lacking.}
\end{Remark}

Before passing to the proof of Theorem~\ref{Tpos}, we note the
following variation thereof, where no restrictions on $\trg  K$
are made:
\begin{theo}\label{Tposvac} Suppose that $(M,g)$ is geodesically complete without
boundary. If $(M,g,K)$ contains an end which is $C^4\times C^3$,
or $C^1$ and polyhomogeneously, compactifiable and which is vacuum
near the conformal boundary, then the conclusions of
Theorem~\ref{Tpos} hold.
\end{theo}

\proof It follows from \cite[Eq.~(C.84)]{CJK} and from the results
of Friedrich that one can deform $M$ near $\scrip$ in the maximal
globally hyperbolic vacuum development of the initial data there
so that the hypotheses of Theorem~\ref{Tpos} hold. The
Trautman-Bondi four-momentum of the original hypersurface
coincides with that of the deformed one
by~\cite[Section~6.1]{CJK}. \qed

\medskip

 \noindent{\sc Proof of Theorem~\ref{Tpos}:} The proof follows the usual argument as proposed by
Witten. While the method of proof is well known, there are tedious
technicalities which need to be taken care of to make sure that
the argument applies.

 Let $D$ be the covariant-derivative operator
of the metric $g$, let $\nabla$ be the covariant-derivative
operator of the space-time metric $\fourg$, and let $\dirac$ be
the  Dirac operator associated with $\nabla$ along $M$,
$$\dirac \psi = \gamma^i\nabla_i \psi $$
(summation over space indices only). Recall the
Sen-Witten~\cite{Witten:mass,Sen} identity:
\begin{equation}\label{weitzenbock0}
\int_{M\setminus\{r\ge R\}} \|\nabla \psi  \|^2_g +\la \psi, (\rho
+ J^i\gamma_i\gamma_0)\psi\ra - \|{\dirac}\psi \|^2_g = \int_{S_R}
B^i(\psi )dS_i\;,
\end{equation}
where the boundary integrand  is
\begin{equation} B^i(\psi) = \la {\nabla}^i \psi
+ \gamma^i{\dirac}\psi , \psi \ra_g\;. \end{equation} We have the
following:

\begin{Lemma}
\label{Lpos1} Let $\psi$ be a Killing spinor for the space-time
background metric $b$, and let $A^\mu\partial_\mu$ be  the
associated translational Killing vector field $A^\mu=\psi^\dagger
\gamma^\mu \psi$. We have
\bel{positiv} \lim_{R\to\infty}\int_{S(R)} < \psi, \nabla^i \psi+
\gamma^i\gamma^j \nabla_j\psi> dS_i =\frac 1{4\pi}p_\mu A^\mu \; .
\ee
\end{Lemma}

\proof  For $A^\mu\partial_\mu =
\partial_0$ the
calculations are carried out in detail in Appendix~\ref{APLpos1}.
For general $A^\mu$ the result follows then by the well known
Lorentz-covariance of $p_\mu$ under changes of Bondi frames (see
{\em e.g.\/}~\cite[Section~6.8]{CJK} for a proof under the current
asymptotic conditions). \qed

\begin{Lemma}
\label{Lpos2} Let $\psi$ be a spinor field on $M$ which vanishes
outside of $\Mext$, and coincides with a Killing spinor for the
background metric $b$ for $R$ large enough. Then
$$\nabla \psi \in L^2(M)\;.$$
\end{Lemma}

\proof In Appendix~\ref{APLpos2} we show that \bel{bobo1} \dirac
\psi \in L^2\;.\ee We then rewrite \eq{weitzenbock0} as
\begin{equation}\label{lichn?}
\int_{M\setminus\{r\ge R\}} \|\nabla \psi  \|^2_g+\la \psi, (\rho
+ J^i\gamma_i\gamma_0)\psi\ra  = \int_{M\setminus\{r\ge R\}}
\|{\dirac}\psi \|^2_g + \int_{S_R} B^i(\psi )dS_i\;.
\end{equation}
By the dominant energy condition \eq{dec} the function $\la \psi,
(\rho + J^i\gamma_i)\psi\ra$ is non-negative. Passing with $R$ to
infinity the right-hand-side is bounded by \eq{bobo1}  and
by Lemma~\ref{Lpos1}. The result follows from the monotone convergence
theorem.\qed

Lemmata~\ref{Lpos1} and \ref{Lpos2} are the two elements which are
needed to establish positivity of the right-hand-side of
\eq{positiv} whatever $A^\mu$, see {\em
e.g.\/}~\cite{ChHerzlich,AndDahl} for a detailed exposition in a
related setting,
or~\cite{bartnik:mass,ParkerTaubes82,ChBlesHouches} in the
asymptotically flat context. (For instance, Lemma~\ref{Lpos2}
justifies the right-hand-side of the
implication~\cite[Equation~(4.17)]{ChHerzlich}, while
Lemma~\ref{Lpos1} replaces all the calculations that
follow~\cite[Equation~(4.18)]{ChHerzlich}. The remaining arguments
in~\cite{ChHerzlich} require only trivial modifications.)
\qed

One also has a version of Theorem~\ref{Tpos} with trapped
boundary, using solutions of the Dirac equation with the boundary
conditions of~\cite{GHHP83} (compare~\cite{Herzlich:mass}):

\begin{theorem} \label{Tposb} Let $(M,g)$ be a
geodesically complete manifold with compact boundary $\partial M$,
and assume that the remaining hypotheses of Theorem~\ref{Tpos} or
of Theorem~\ref{Tposvac} hold. If $\partial M$ is either
outwards-past trapped, or outwards-future trapped, then $p^\mu$ is
{\em timelike future directed}:
$$p^0 > \sqrt{\sum_i p_i^2}\;.$$

\qed
\end{theorem}

\section{The mass of approximate Bondi foliations}
\label{sec:3}

 The main tool
in our analysis so far was the foliation of the asymptotic region
by spheres arising from space-times Bondi coordinates adapted to
the initial data surface. Such foliations will be called
\emph{Bondi foliations}. The aim of this section is to reformulate
our definition of mass as an object directly associated to this
foliation. The definition below is somewhat similar to that of
Brown, Lau and York~\cite{LauYorkBrown}, but the normalisation
(before passing to the limit) used here seems to be different from
the one used by those authors.

Let us start by introducing some notation  (for ease of reference
we collect here all notations, including some which have already
been introduced elsewhere in the paper): Let \( \fourg_{\mu \nu }
\) be a metric of a spacetime which is asymptotically flat in null
directions. Let \( \gmetric _{ij} \) be the metric induced on a
three-dimensional surface \( \hyp  \) with extrinsic curvature \(
K_{ij} \).  We denote by \( \Gtrzy ^i{_{jk}} \) the Christoffel
symbols of \( \gmetric _{ij}. \)

We suppose that we have a function $x>0$ the level sets \(
x=\mathrm{const}. \) of which provide a foliation of \( \hyp  \)
by two-dimensional submanifolds, denoted by $\hyp_x$, each of them
homeomorphic to a sphere. The level set $``\{x=0\}"$ (which does
not exist in $\hyp$) should be thought of as corresponding to
$\scrip$. The metric \( \gmetric _{ij} \) induces a metric \( \gtw
_{AB} \) on each of these spheres, with area element
$\lambda=\sqrt{\det \gtw _{AB}}$. By \( k_{AB} \) we denote the
extrinsic curvature of the leaves of the $x$-foliation, \(
k_{AB}={\Gtrzy ^{x}{_{AB}}}/{\sqrt{\gmetric ^{xx}}}, \) with mean
curvature \( k=\gtw ^{AB}k_{AB}. \) (The reader is warned that in
this convention the outwards extrinsic curvature of a sphere in a
flat Riemannian metric is negative.) There are also the
corresponding objects for the background (Minkowski) metric \(
\fourb_{\mu \nu } \), denoted by letters with a circle.

Let \( X \) be a field of space-time vectors defined along \( \hyp
. \) Such a field can be decomposed as \( X=Y+Vn \), where \( Y \)
is a field tangent to \( \hyp , \) \( n \) --- the unit (\(
n^{2}=-1 \)), future-directed vector normal to \( \hyp \).
Motivated by \eq{masssphere}, we define the following functional
depending upon various objects defined on the hypersurface \( \hyp
\) and on a vector field \( X \):
\begin{equation}
\label{int2} \frac{1}{16\pi }\lim _{x\to 0 }\int _{\hyp _{x}}\funk
(X)\rd^{2}x\;,
\end{equation}
where
\bel{int2.1}
\funk (X)=2\sqrt{\det \gmetric}(\zP^{x}{_i}-P^{x}{_i})Y^{i}
+2(\zlambda \zk-\lambda k)V.\ee So far the considerations were
rather general, from now on we assume that  the initial data set
contains an {\em asymptotically CMC} conformally compactifiable
end and that $x$ provides an approximate Bondi foliation, as
defined in Section~\ref{STBah}. (To avoid ambiguities, we
emphasise that we do impose the restrictive condition \eq{cm2hb},
which in terms of the function $\alpha$ of \eq{a1} translates into
\Eq{acond}.) We then have $\lambda=\zlambda$ up to terms which do
not affect the limit $x\to 0$ (see Appendix~\ref{app:obiekty3d})
so that the terms containing $\lambda$ above can also be written
as $\zlambda (\zk- k)$ or $\lambda (\zk- k)$.  We will show that
in the limit \( x\to 0 \) the functional \eq{int2} will give the
Trautman-Bondi mass and momentum, as well as angular momentum and
centre of mass, for appropriately chosen fields \( X \)
corresponding to the relevant generators of Poincar\'e group.

To study the convergence of the functional when  \( x\to 0 \) we
need to calculate several objects on $\hyp$. We write the
spacetime metric in Bondi-Sachs coordinates, as in \eq{gB}  and
use the standard expansions for the coefficients of the metric
(see \emph{eg.}\/~\cite[Equations~(5.98)-(5.101)]{CJK}). The
covariant derivative operator associated with the metric
$\hst_{AB}$ is denoted by ${}_{||A}$.
 Some intermediate
results needed in those calculations are presented in Appendix
\ref{app:obiekty3d}, full details of the calculations can be
found\footnote{The reader is warned that $\funk$ there is $-\funk$
here.} in~\cite{Leski}. Using formulae (\ref{lgg})-(\ref{kk0}) and
the decomposition of Minkowski spacetime Killing vectors given in
Appendix~\ref{app:killingi} we get (both here and in
Appendix~\ref{app:killingi} all indices are coordinate ones):
\[ \funk (X_{time})=\sqrt{\hst
}[4M-\chi^{CD}{_{||CD}}+O(x)]\;,\]
\[
\funk (X_{trans})=-v\sqrt{\hst }[4M-\chi^{CD}{_{||CD}}+O(x)]\;,\]
\begin{eqnarray*}
\funk (X_{rot}) & = &- \sqrt{\hst }\left[\varepsilon ^{AB}
\left(\underline{-\frac{\chi^{C}{_{A||C}}}{x}}+\frac{3(\chi
^{CD}\chi _{CD})_{||A}}{16}
\right.\right.\\
 &  & \left.\left.{}+6N_{A}+\frac{1}{2}\chi _{AC}\chi^{CD}{_{||D}}
 \right)v_{,B}+O(x)\right]\;,
\end{eqnarray*}
\begin{eqnarray*}
\funk (X_{boost}) & = &- \sqrt{\hst }\left[
\underline{-\frac{\chi^{C}{_{A||C}}v^{,A}}{x}}+\frac18 \chi ^{CD}\chi _{CD} v
\right.\\
 &  & {}+\left.\left(\frac{3}{16}(\chi ^{CD}\chi _{CD})_{||A}+6N_{A}
 +\frac{1}{2}\chi _{AC}\chi^{CD}{_{||D}}\right)v^{,A}+O(x)\right]\;.\nonumber
\end{eqnarray*} The underlined terms in integrands
corresponding to boosts and rotations diverge when $x$ tends to
zero, but they yield zero when integrated over a sphere. In the
formulae above we set \(u-\alpha\) equal to any constant, except
for the last one where \(u-\alpha=0\). Moreover, \( v \) is a
function on the sphere which is a linear combination of $\ell=1$
spherical harmonics and \( \varepsilon ^{AB} \) is an
antisymmetric tensor (more precise definitions are given in
Appendix \ref{app:killingi}).

In particular we obtain: \begin{equation} \label{etb}
E_{TB}=\frac{1}{16\pi }\lim _{x\to 0}\int _{\hyp _{x}}\funk
(X_{time})\rd^{2}x\;,
\end{equation}
 \begin{equation}
\label{ptb} P_{TB}=\frac{1}{16\pi }\lim _{x\to 0}\int _{\partial
\hyp _{R}}\funk (X_{trans})\rd^{2}x\;,
\end{equation}
where the momentum is computed for a space-translation generator
corresponding to the function \( v \) (see Appendix
\ref{app:killingi}). It follows that the integrals of \( \funk (X)
\) are convergent for all four families of  fields $X$.

\subsection{Polyhomogeneous metrics}
\label{sec:log}

In this section we consider polyhomogeneous metrics. More
precisely, we will consider  metrics  of the form \eq{gB0} for
which the \( V, \) \( \beta , \) \( U^{A} \) and \( h_{AB} \) have
asymptotic expansions of the form\[ f\simeq \sum _{i}\sum
_{j=1}^{N_{i}}f_{ij}(x^{A})x^{i}\log ^{j}x\;,\] where the
coefficients \( f_{ij} \) are smooth functions. This means that \(
f \)  can be approximated up to terms \( O(x^{N}) \) (for any \( N
\)) by a finite sum of terms of the form \( f_{ij}(x^{A})x^{i}\log
^{j}x, \) and that this property is preserved under
differentiation in the obvious way.

As mentioned in~\cite[Section~6.10]{CJK}, allowing a
polyhomogeneous expansion of \( h_{AB} \) of the form \[
h_{AB}=\hst _{AB}+x\chi _{AB}+x\log xD_{AB}+o(x)\] is not
compatible with the Hamiltonian approach presented there because
the integral defining the symplectic structure diverges. (For such
metrics it is still possible to define the Trautman-Bondi mass and
the momentum~\cite{CJM}.) It is also noticed
in~\cite[Section~6.10]{CJK}, that when logarithmic terms in \(
h_{AB} \) are allowed only at the \( x^{3} \) level and higher,
then all integrals of interest converge. It is therefore natural
to study metrics for which \( h_{AB} \) has the intermediate
behavior
\begin{equation}
\label{log} h_{AB}=\hst _{AB}+x\chi _{AB}+x^{2}\zeta_{AB}(\log
x)+O(x^{2})\;,
\end{equation}
where \( \zeta_{AB}(\log x) \) is a polynomial of order \( N \) in
\( \log x \) with coefficients being smooth, symmetric tensor
fields on the sphere. Under \eq{log}, for the $\funk$ functional
we find:\[ \funk (X_{time})=\sqrt{\hst
}[4M-\chi^{CD}{_{||CD}}+O(x\log ^{N+1}x)]\;,\]
\[
\funk (X_{trans})=-v\sqrt{\hst }[4M-\chi^{CD}{_{||CD}}+O(x\log
^{N+1}x)]\;.\] The only difference with the power-series case is
that terms with \( O(x\log ^{N+1}x) \) asymptotic behavior appear
in the error term, while previously we had \( O(x). \) Thus
$\funk$ reproduces the Trautman-Bondi four-momentum for such
metrics as well.

In \cite{CJK} no detailed analysis of the corresponding
expressions for the remaining generators ({\em i.e.}\/, \( X_{rot}
\) and \( X_{boost} \)) was carried out, except for a remark in
Section 6.10, that the asymptotic behavior (\ref{log}) leads to
potentially divergent terms in the Freud potential, which might
lead to a logarithmic divergence for boosts and rotations in the
relevant Hamiltonians, which then could cease to be well defined.
Here we check that the values \( \funk (X_{rot}) \) and \( \funk
(X_{boost}) \) remain well defined for certain metrics of the form
(\ref{log}). Supposing that \( h_{AB} \) is of such form one finds
(see \cite[Appendix C.1.2]{CJK})

\[
\partial _{u}\zeta_{AB}=0\;,\]
 \[
U^{A}=-\frac{1}{2}x^{2}\chi^{AC}{_{||C}}+x^3 W^{A}(\log
x)+O(x)\;,\]
 \[
\beta =-\frac{1}{32}x^{2}\chi ^{CD}\chi _{CD}+O(x^{3}\log ^{2N}x)\;,\]
 \[
V=\frac{1}{x}-2M+O(x\log ^{N+1}x)\;,\]
 where \( W^{A}(\log x) \) is a polynomial of order \( N+1 \)
in \( \log x \) with coefficients being smooth vector fields on
the sphere. The \( W^{A} \)'s can be calculated by solving
Einstein equations (which are presented in convenient form in
\cite{ChMS}).

The calculation in case of polyhomogeneous metrics is very similar
to the one in case of metrics allowing power-series expansion (see
Appendix \ref{app:obiekty3dlog}), leading to
\[
\funk (X_{rot})=-\sqrt{\hst }\left[\varepsilon ^{AB}
\left(\underline{-\frac{\chi^{C}{_{A||C}}}{x}}-3W_{A}-xW_{A\uder
x}\right)v_{,B}+O(1)\right]\;,\]
\[
\funk (X_{boost})=-\sqrt{\hst
}\left[\left(\underline{-\frac{\chi^{C}{_{A||C}}}{x}}
-3W_{A}-xW_{A\uder x}\right)v^{,A}+O(1)\right]\;,\] where
${}_{\uder x}$ denotes derivation at constant $u$. As in the
previous section, the underlined terms tend to infinity for \(
x\to 0, \) but  give zero when integrated over the sphere. However
some potentially divergent terms remain, which can be rewritten as
\[ \frac{1}{x^{2}}\varepsilon ^{AB}(x^{3}W_{A})_{\uder
x}v_{,B}\;,\qquad \frac{1}{x^{2}}(x^{3}W_{A})_{\uder x}v^{,A}\;.\]
To obtain convergence one must therefore have:\[ \lim _{x\to
0}\int _{\hyp _{x}}\frac{1}{x^{2}}\varepsilon ^{AB}
(x^{3}W_{A})_{\uder x}v_{,B}\sqrt{\hst }\rd^{2}x=0\] and\[ \lim
_{x\to 0 }\int _{\hyp _{x}}\frac{1}{x^{2}} (x^{3}W_{A})_{\uder
x}v^{,A}\sqrt{\hst }\rd^{2}x=0\] for any function $v$ which is a
linear combination of the $\ell=1$ spherical
harmonics.\footnote{See Appendix \ref{app:killingi}. }
It turns out that in the simplest case $ N=0 $ those integrals are
identically zero when vacuum Einstein equations are
imposed~\cite{ChMS}.

\subsection{The Hawking mass of approximate Bondi spheres}
 One of the
objects of interest associated to the two dimensional surfaces
$\hyp_x$ is their Hawking mass,
\begin{eqnarray} \nonumber m_H(\hyp_x) & = & \sqrt{\frac A{16\pi}}\left(1- \frac 1
{16\pi} \int_{\hyp_x} \theta^-\theta^+ d^2\mu\right) \\ & = &
\sqrt{\frac A{16\pi}}\left(1+ \frac 1 {16\pi} \int_{\hyp_x}
\lambda \left(\frac {P^{xx}}{g^{xx}}- k\right) \left(\frac
{P^{xx}}{g^{xx}}+ k\right) d^2 x\right)
 . \label{Hm}
\end{eqnarray} We wish to show that for asymptotically CMC
initial data the Hawking mass of approximate Bondi spheres
converges to the Trautman Bondi mass. In fact, for $a=(a_\mu)$ let
us set
$$v(a)(\theta,\varphi) = a_0 + a_i \frac{x^i}{r}\;,$$
where $x^i$ has to be expressed in terms of the spherical
coordinates in the usual way, and
\beaa
p_H(a,\hyp_x) & = & p^\mu_Ha_\mu \\ & = & -\frac 1
{16\pi}\sqrt{\frac A{16\pi}}\int_{\hyp_x} v(a)\left(\frac{16\pi}{
A}- \theta^-\theta^+ \right)d^2\mu\\ & = & -\frac 1
{16\pi}\sqrt{\frac A{16\pi}}\int_{\hyp_x} v(a)\left(\frac{16\pi}{
A}+ \left(\frac {P^{xx}}{g^{xx}}\right)^2-  k^2 \right)d^2\mu\;,
\eeaa with $d^2\mu=\lambda d^2x$. Up to the order needed to
calculate the limit of the integral, on approximate Bondi spheres
satisfying \eq{acond} we have (see Appendices~\ref{app:obiekty3d}
and \ref{app:obiekty3dlog})
\[
\frac {P^{xx}}{g^{xx}}  =  P^x{_x} + O(x^4)\;, \quad \lambda =
\frac{\sqrt{\hst}}{x^2}+O(x^3)\;,
\]
and
\beaa
\Big(\frac {P^{xx}}{g^{xx}}\Big)^2 & = & \frac{2}{\alpha_{,x}} -
\frac{4\beta}{\alpha_{,x}} - 3x^2 (1-2Mx)
- x^3 \chi^{CD}{_{||CD}} + \Ol(x^4)\;,\\
k^2 & = & \frac{2}{\alpha_{,x}} - \frac{4\beta}{\alpha_{,x}} + x^2
(1-2Mx) + x^3 \chi^{CD}{_{||CD}} + \Ol(x^4)\;. \eeaa Therefore
\[
\lambda\Bigg(\Big(\frac {P^{xx}}{g^{xx}}\Big)^2-k^2\Bigg) = -4
\sqrt{\hst} + \sqrt{\hst}[2x(4M-\chi^{CD}{_{||CD}})+\Ol(x^2)]\;,\]
which, together with
\[A = \frac{4 \pi}{x^2} +O(x^3)\;,\quad
\sqrt\frac{A}{16\pi} = \frac{1}{2x} + O(x^4)\;,\] yields
\[ p_H(a,\hyp_x )= -\frac{1}{16\pi} \int_{\hyp_x}  v(a)[\sqrt{\hst}
(4M - \chi^{CD}{_{||CD}}) + \Ol(x)] \rd^{2} x \;.\] Passing to the
limit $x\to 0$ the $\chi^{CD}{_{||CD}}$ term integrates out to
zero, leading to equality of the Trautman-Bondi four-momentum
$p_{\mbox{\scriptsize \rm TB}}(a,\hyp)$ with the limit of
$p_H(a,\hyp_x )$ defined above.

\section{Acknowledgements}
This research was supported in part by the Polish Research Council
grant KBN 2 P03B 073 24 and by the Erwin Schr\"odinger Institute.
PTC acknowledges a travel grant from the Vienna City Council. SzL
was also supported by a scholarship from the Foundation for Polish
Science.
\appendix

\section{Proof of \eq{eq:3.2n}}
\label{AVpot}

From
\begin{equation}\label{divYPg}
  2 D_l (Y^k P^l{_k}) = P^{kl}\zA_{kl} + Y^k J_k(g,K) +
  P^{kl} \left( 2V\zK_{kl}+\lie_Y e_{kl} \right)
\end{equation}
we get
\begin{eqnarray}\nonumber
  V R_g  + 2 D_l (Y^k P^l{_k}) & =&
 V \rho(g,K)+ Y^k J_k(g,K) + P^{kl}\zA_{kl}
 \\\nonumber  &&
  - V \left(\zP^{mn}\zP_{mn}-\frac1{n-1} (\trb \zP)^2  \right) + \zP^{kl}\lie_Y e_{kl} \\
    & & + V e_{kl} \left(2\zP^{ml}\zP_m{_k}- \frac2{n-1}\zP^{kl}\trb \zP \right)
     + Q_2
 \;,\nonumber \\ &&\label{VRYP} \end{eqnarray}
where, here and below, we use the symbol $Q_i$ to denote terms
which are quadratic or higher order in $e$ and $P-\zP$. Moreover,
we have
\begin{eqnarray}\label{Ple}
  \zP^{kl}\lie_Y e_{kl} &=& -e_{kl} \lie_Y \zP^{kl} +
  D_m(Y^m \zP^{kl}e_{kl}) - \zP^{kl}e_{kl} D_m Y^m \; ,\\
 D_m Y^m &=& \frac12g^{kl}(\zA_{kl} + \lie_Y e_{kl})+
 V(g^{kl}-b^{kl})\zK_{kl} + V\trg \zK \nonumber\\ &=:&
 \frac12g^{kl} \zA_{kl} + \frac V{n-1} \trb\zP + L_1\;,\label{DY}
 \\\nonumber
  2 D_l (Y^k \zP^l{_k}) & = & \zP^{kl}\zA_{kl} + Y^k J_k(b,\zK)
   -2V\left( \zP^{mn}\zP_{mn}-\frac1{n-1}(\trb \zP)^2   \right) \\ & &
   + D_l \left(Y^k \zP^l{_k}b^{mn}e_{mn}\right)
   - b^{mn}e_{mn}\zD_l (Y^k \zP^l{_k}) + Q_1 \; .
 \label{DYzP}
\end{eqnarray}
(The term $L_1$ in \eq{DY} is  a linear remainder term which,
however, will give a quadratic contribution  in equations such as
\eq{divY2} below.) Using
\begin{eqnarray}\nonumber
 -\zD_l (Y^k \zP^l{_k}) &= & V\left(\zP^{mn}\zP_{mn}-\frac1{n-1} (\trb \zP)^2  \right)
 -\frac12 \zP^{kl}\zA_{kl}-\frac12 Y^k J_k(b,\zK) \\
 & = & V R_b - V\rho(b,\zK) -\frac12 \zP^{kl}\zA_{kl}-\frac12 Y^k J_k(b,\zK)
 \label{divYP}
\end{eqnarray}
and \eq{VRYP}-\eq{DYzP}
 we are led
to
\begin{eqnarray} \nonumber
  V(R_g-R_b) & = & V\left(\rho(g,K)-\rho(b,\zK)\right)
  + Y^k\left( J_{k}(g,K)-J_k(b,\zK)\right)
  - \frac {\partial_l {\mathbb V}^l}{\sqrt{\det g}}   \\
   \nonumber
  & & \hspace*{-1.7cm} + V \left[ 2\zP^{ml}\zP_m{}^{k}
  -\frac3{n-1}\trb \zP\zP^{kl} -
  \left( \zP^{mn}\zP_{mn}-\frac1{n-1}(\trb \zP)^2  \right)b^{kl}
  \right] e_{kl} \\& & \hspace*{-1cm}
    - e_{kl}\lie_Y \zP^{kl} +\frac12 e_{kl}
  \left[ b^{kl}\left(\zP^{mn}\zA_{mn}+Y^m J_m(b,\zK) \right)
  -b^{mn}\zA_{mn}\zP^{kl}\right]\nonumber
   \\& & + (P^{kl} -\zP^{kl})\zA_{kl} + Q_3 \;,
   \label{divY2}
\end{eqnarray}
where
  \bel{matV} {\mathbb V}^l(Y):=  2\sqrt{\det g}\left[ (P^l{_k}
-\zP^l{_k})Y^k -\frac12
  Y^l\zP^{mn}e_{mn} +\frac12Y^k \zP^l{_k}b^{mn}e_{mn}  \right] \;.
  \ee
  Inserting
\Eq{eq:3.2} into \eq{divY2} one obtains the following counterpart
of \Eq{eq:3.2}
\begin{eqnarray}
 \partial_i \left(\ourU^i(V)+ \ourV^i(Y)\right)& =&\sqrt{\det g}
\;\left[ V\left(\rho(g,K)-\rho(b,\zK)\right)\right.\nonumber
\\ &&\label{eq:3.2n1}\left.
 + Y^k\sqrt{\det g}\left(
J_{k}(g,K)-J_k(b,\zK)\right)+ s' + Q''\right]\;,\nonumber \\ &&
\end{eqnarray}
where $Q''$ contains terms which are quadratic in the deviation of
$g$ from $b$ and its derivatives, and in the deviations of $K$
from $\mathring K$, while $s'$, obtained by collecting all terms
linear or linearised in $e_{ij}$, except for those involving
$\rho$ and $J$, reads
$$s'= (\zS^{kl}+\zB^{kl}) e_{kl}+(P^{kl}-\zP^{kl})\zA_{kl}\;,$$
\begin{equation}\label{linw1}
\zB^{kl}:= \frac12
  \left[ b^{kl}\zP^{mn}\zA_{mn}
  -b^{mn}\zA_{mn}\zP^{kl}\right]
\;.
\end{equation}

\section{ $3+1$ charge integrals \emph{vs}  the Freud integrals}
\label{Slin} In this appendix we wish to show that under the
boundary conditions of Section~\ref{sec:rd} the numerical value of
the Riemannian charge integrals coincides with that of the
Hamiltonians derived in a space-time setting in
\cite{ChAIHP,ChNagy}:
\begin{eqnarray}
  H^\mu & \equiv & \KA^{\mu}_{\alpha\beta}\lie_X
\Kp^{\alpha\beta} -X^\mu L =
\partial_\alpha
\ourW^{\mu\alpha} + \frac 1 {8\pi}\mcG^\mu{}_\beta(\Lambda)
X^\beta\;,
  \label{Fsup1}
\\
 \ourW^{\nu\lambda}&= &
{\ourW^{\nu\lambda}}_{\beta}X^\beta - \frac 1{8\pi} \sqrt{|\det
g_{\rho\sigma}|} g^{\alpha[\nu}\delta^{\lambda]}_\beta
{X^\beta}_{;\alpha} \;,\label{Fsup2new}
\\ {\ourW^{\nu\lambda}}_\beta &= & \displaystyle{\frac{2|\det
  \bmetric_{\mu\nu}|}{ 16\pi\sqrt{|\det g_{\rho\sigma}|}}}
g_{\beta\gamma}\left(\frac{{|\det
    g_{\rho\sigma}|}}{{|\det\bmetric_{\mu\nu}|}} g^{\gamma[\lambda}g^{\nu]\kappa}\right)_{;\kappa}
\nonumber \\  &= & 2
\Kp_{\beta\gamma}\left(\Kp^{\gamma[\lambda}\Kp^{\nu]\kappa}\right)_{;\kappa}
\nonumber
\\ & = & 2 \Kp^{\mu[\nu}\KA^{\lambda]}_{\mu\beta} -2
\delta^{[\nu}_\beta \KA^{\lambda]}_{\mu\sigma}\Kp^{\mu\sigma} -
\frac 2 3 \Kp^{\mu[\nu}\delta^{\lambda]}_\beta
\KA^{\sigma}_{\mu\sigma} \;,\label{Freud2.0}
\end{eqnarray}
where a semi-colon denotes the covariant derivative of the metric
$\bmetric$, square brackets denote anti-symmetrisation (with a
factor of 1/2 when two indices are involved), as before
$\Kp_{\beta\gamma}\equiv (\Kp^{\alpha\sigma})^{-1}= 16\pi
g_{\beta\gamma}/\sqrt{|\det g_{\rho\sigma}|}$.
    Further, $\mcG^\alpha{}_\beta(\Lambda)$ is the Einstein tensor density eventually shifted by a cosmological constant,
\bel{ETD}\mcG^\alpha{}_\beta(\Lambda):= \sqrt{|\det g_{\rho\sigma}|}
\left(R^\alpha{}_\beta - \frac 12 g^{\mu\nu}R_{\mu\nu}
    \delta^\alpha_\beta + \Lambda \delta ^\alpha_\beta\right) \ee
    (equal, of course, to the energy-momentum
    tensor density
    of the matter fields in models with matter, and vanishing in vacuum), while
\begin{eqnarray}\label{Pend2}
  \KA^{\lambda}_{\mu\nu} & = &  \frac 12
  \Kp_{\mu\alpha}  \Kp^{\lambda\alpha}_{\ \ ; \nu}
  + \frac 12
  \Kp_{\nu\alpha}  \Kp^{\lambda\alpha}_{\ \ ; \mu}
  - \frac 12
  \Kp^{\lambda\alpha} \Kp_{\sigma\mu} \Kp_{\rho\nu}
  \Kp^{\sigma\rho}_{\ \ ; \alpha}
  \nonumber \\   & &
  +\frac 14
  \Kp^{\lambda\alpha} \Kp_{\mu\nu} \Kp_{\sigma\rho}
  \Kp^{\sigma\rho}_{\ \ ; \alpha} \;,
\end{eqnarray}
where by $\Kp_{\mu\nu}$ we denote the matrix inverse to
$\Kp^{\mu\nu}$, and \bel{Ldef} L := \Kp^{\mu\nu} \left[ \left(
    {\Gamma}^{\alpha}_{\sigma\mu} - {\Bgamma}^{\alpha}_{\sigma\mu}
  \right) \left( {\Gamma}^{\sigma}_{\alpha\nu}-
    {\Bgamma}^{\sigma}_{\alpha\nu} \right) - \left(
    {\Gamma}^{\alpha}_{\mu\nu} - {\Bgamma}^{\alpha}_{\mu\nu} \right)
  \left( {\Gamma}^{\sigma}_{\alpha\sigma} -
    {\Bgamma}^{\sigma}_{\alpha\sigma} \right) + r_{\mu\nu} \right] \;.
\ee Finally, $B^\alpha_{\mu\nu}$ is the connection of the
background metric, and $r_{\mu\nu}$ its Ricci tensor -- zero in
our case.  For typesetting reasons in the remainder of this
appendix we write $\eta$ for $\fourb$, while the symbol $b$ will
be reserved for the space-part of $\fourb$ and its inverse. While
$\eta$ is flat, the reader should not assume that it takes the
usual diagonal form. Let us denote by $\approx$ the linearisation
at $\fourb\equiv \eta$; we find
\begin{equation} 16\pi {\ourW^{0l}}_0 \approx \sqrt{-\eta}\left(
e^l{_m}{^{;m}} - e^m{_m}{^{;l}}\right)\;,\label{b7}
\end{equation}
\begin{equation}\label{b8}
16\pi {\ourW^{0l}}_k \approx \sqrt{-\eta}\left[
\delta^l{_k}\left(e^m{_m}{^{;0}} - e^0{_m}{^{;m}} \right) +
e^0{_k}{^{;l}} - e^l{_k}{^{;0}}\right]\;,
\end{equation}
leading to
\begin{equation}
16\pi {\ourW^{0l}}_\beta n^{\beta} \approx \sqrt{b} \left(
b^{kn}b^{lm}-b^{km}b^{ln} \right) e_{mk;n}\;,
\end{equation}
\begin{equation}
16\pi
{\ourW^{0l}}_k Y^k \approx \sqrt{-\eta}
\left( Y^l b^{km}-Y^k b^{lm}\right)\left( e_{mk}{^{;0}} - e^0{_{k;m}}
 \right)\;.
\end{equation}
Here $$e_{\mu\nu}:= g_{\mu\nu}-\eta_{\mu\nu}\;,$$
and in the
linearised expressions of this appendix all space-time indices are
raised and lowered with $\eta$. Similarly,
\begin{eqnarray} & &
\left[\sqrt{-g}(g^{l\alpha}\eta^{0\mu} - g^{0\alpha}\eta^{l\mu})
- \sqrt{-\eta}(\eta^{l\alpha}\eta^{0\mu} - \eta^{0\alpha}\eta^{l\mu})\right]
X_{\mu;\alpha} \approx \nonumber\\
& &
\sqrt{-\eta}\left[ (\eta^{l\alpha}\eta^{0\mu} - \eta^{0\alpha}\eta^{l\mu})\frac12
e^\sigma{_\sigma}
 - e^{l\alpha}\eta^{0\mu} + e^{0\alpha}\eta^{l\mu}\right]
X_{\mu;\alpha} = \nonumber \\ & & \sqrt{-\eta}\left(
b^{mn}e_{mn}b^{lk}X^0{}_{;k} - b^{ln}b^{mk}e_{mn}X^0{}_{;k}
 + e^0{_m}b^{mk}b^{ln}X_{n;k} \right)\;.
\end{eqnarray} We also have
\begin{equation}
X^0{}_{;k}= \frac1{\zN}\left( V_{,k} - Y^l \zK_{lk}\right)\;,
\end{equation}
\begin{equation}
X_{l;k}= b_{lm}\zD_k Y^m - V \zK_{lk}\;,
\end{equation}
and we have of course assumed that \[ X=Vn+Y \] is a background
Killing vector field, with $Y$ tangent to the hypersurface of
interest,  so that
\begin{equation}
 b_{lm}\zD_k Y^m + b_{km}\zD_l Y^m = 2V\zK_{lk}\;.
 \end{equation}
We recall that\[
n^{\mu}=-\frac{\eta^{0\mu}}{\sqrt{-\eta^{00}}}\;,\] which gives
$X^0=\frac V{\zN}$, $X^k=Y^k-\frac V{\zN}\zN^k$, with the lapse
and shift given by the formulae
 $\zN=\frac1{\sqrt{-\eta^{00}}}$,
 $\zN^k=-\frac{\eta^{0k}}{\eta^{00}}$.
 We will also need the $3+1$ decomposition of the Christoffel
 symbols $B^\alpha_{\beta\gamma}$ of the four-dimensional background metric in terms of those,
 denoted by
  $\zGamma^m{_{kl}}(b)$,
 associated with the three-dimensional one:
 \[ B^m{_{kl}}= \zGamma^m{_{kl}}(b) +
 \frac{\zN^m}{\zN}\zK_{lk}\;,\quad B^0{_{0k}}=\partial_k \log\zN - \frac{\zN^l}{\zN}\zK_{lk}\;,
\]
 \[  B^0{_{kl}}= - \frac{1}{\zN}\zK_{lk} \;,\quad B^l{_{k0}}= \zD_k\zN^l - \frac{\zN^l}{\zN}\zD_k\zN
  -\zN \zK^l{_k} +\frac{\zN^l}{\zN}\zN^m\zK_{mk} \; . \]
Linearising the equation for $K_{ij}$ one finds \begin{equation}
 \delta K_{kl}:= -\frac12 \zN \left( e^0{_{k;l}} + e^0{_{l;k}} - e_{kl}{^{;0}} \right)
 -\frac12 (\zN)^2 e^{00}\zK_{kl}\;,
 \end{equation}
 where the relevant indices have been raised with the space-time background
 metric.
 This leads to the following formula for the linearised ADM
 momentum:
\begin{equation}
 \delta P^l{_k}:=\delta^l{_k} b^{mn} \delta K_{mn} -  b^{ml} \delta K_{mk}
  +\left( b^{li} \zK^j{_k} -\delta^l{_k} \zK^{ij}\right)e_{ij}\;.
 \end{equation}
 A rather lengthy calculation leads then to
\begin{eqnarray}
& & \nonumber \left[\sqrt{-g}(g^{l\alpha}\eta^{0\mu} -
g^{0\alpha}\eta^{l\mu}) - \sqrt{-\eta}(\eta^{l\alpha}\eta^{0\mu} -
\eta^{0\alpha}\eta^{l\mu})\right] X_{\mu;\alpha}  \\\nonumber & &
+  16\pi V{\ourW^{0l}}_\beta n^{\beta} + 16\pi{\ourW^{0l}}_k Y^k
 \approx  \\ & & \nonumber
\sqrt{b} V b^{ij}b^{lm}(\zD_i e_{jm} - \zD_m e_{ji}) + \sqrt{b}
e_{mn}(b^{mn}b^{lk} -b^{ln}b^{mk})\zD_k V  \\ & & \nonumber
+\sqrt{b}\zD_k \left[ \zN e^0{_m}(Y^l b^{km} -Y^k b^{lm})\right] +
2\sqrt{b}(Y^l b^{km} -Y^k b^{lm})\delta K_{km} \\\nonumber & & +
\sqrt{b}e_{mn} \left[ 2Y^k\zK_k{^m}b^{ln}
 -Y^k\zK_k{^l}b^{mn} - Y^l \zK^{mn} \right] = \\ & & \nonumber
\sqrt{b} V b^{ij}b^{lm}(\zD_i e_{jm} - \zD_m e_{ji}) +
\sqrt{b} e_{mn}(b^{mn}b^{lk} -b^{ln}b^{mk})\zD_k V  \\ & & \nonumber
+\partial_k \left[ \sqrt{b}\zN e^0{_m}(Y^l b^{km} -Y^k b^{lm})\right]
 + 2 \sqrt{b} Y^k \delta P^l{_k} \\ & &
 + \sqrt{b} e_{mn} \left( \zP^l{_k}Y^k b^{mn} - Y^l\zP^{mn}
 \right)\;.\label{linnon}
\end{eqnarray}
The integral over a sphere of \eq{linnon} coincides with the
linearisation of the integral over a sphere of $\ourU^i+\ourV^i$,
as the difference between those expressions is a complete
divergence.

 Let us finally show that the linearised
expression above does \emph{not} reproduce the Trautman-Bondi
mass. It is most convenient to work directly in space-time Bondi
coordinates $(u,x,x^A)$ rather than in coordinates adapted to
$\hyp$. We consider a metric of the Bondi form \eq{gB0}--\eq{gb1}
with the asymptotic behavior \eq{beta}--\eq{hU}; we have
\[ e_{00}=1-xV+x^{-2}h_{AB}U^AU^B = 2Mx + O_{\ln^* x}(x^2)\;,\]
\[ e_{03}=x^{-2}(e^{2\beta}-1) = O(1)\;,\]
\[ e_{0A}=-x^{-2} h_{AB}U^B = \frac12 \chi_A{^B}{_{||B}}+O(x)\;,\]
\[ e_{AB}= x^{-2}(h_{AB}-\hst_{AB})= x^{-1}\chi_{AB} + O(1)\;, \]
\[ e_{33}=e_{3A}=0\;. \]
(Recall that $\hst$ denotes the standard round metric on $S^2$.)
For translational $\fourb$-Killing vector fields $X^\mu$ we have
$X_{\mu;\nu}=0$, hence
$$ \ourW^{0l} =  \ourW^{0l}{_\mu}X^\mu \, .$$
Now, taking into account \eq{b7} and \eq{b8}, the linearised Freud
superpotential takes the form:
\begin{eqnarray}
\nonumber 16\pi \ourW^{03} & = & \sqrt{-\eta} \left[
(e^A{_A}{^{;3}} - e^3{_A}{^{;A}})X^0 + (e^0{_A}{^{;A}} -
e^A{_A}{^{;0}})X^3\right. \\ &&\left. + (e^3{_A}{^{;0}} - e^0{_A}{^{;3}})X^A
 \right] \;, \label{linFreud} \end{eqnarray}
 The following formulae for the non-vanishing
${{}^4\Bgamma^\sigma}_{\beta\gamma}$'s for the flat metric
\eq{backg1} are useful when working out \eq{linFreud}:
\be\label{conn1} {}^4B^\myzero {_{AB}} =x^{-1}\hst_{AB} ,\quad
{}^4B^\three {_{AB}} = x\hst_{AB} ,\quad {}^4B^A{_{\three B}}
=-x^{-1}\delta^A_B , \ee\be\label{conn2} {}^4B^\three {_{\three
\three }} =-2x^{-1} ,\quad {}^4B^A{_{BC}} =\Gst^A{_{BC}}(\hst)\;.
\ee The Killing field corresponding to translations in spacetime
can be characterised by a function $\kappa$ which is a linear
combination of the $\ell=0$ and $\ell=1$ spherical harmonics:
\[ X^0=\kappa \; , \quad X^3=-\frac12 x^2\dtwo\kappa \; , \quad
X^A=-x\hst^{AB}\kappa_{,B} \;.\] With some work one finds the
following formula for the linearised superpotential for time
translations ($\kappa=1$)
\beaa \nonumber 16\pi \ourW^{03}& = &\sqrt{-\eta}(e^A{_A}{^{;3}} - e^3{_A}{^{;A}})
\\ &=&\sin\theta \left[
   4M -\frac12\chi^{AB}{_{||AB}}+\chi^{CD}\partial_u\chi_{CD} +\Ol(x) \right]\;.
  \label{badeq} \eeaa
   The resulting integral reproduces the Trautman-Bondi
   mass if and only if  the $\chi^{CD}\partial_u\chi_{CD}$ term above gives a zero contribution
   after being integrated upon; in general this will not be the case.

\section{Proof of Lemma~\ref{Lpos1}}\label{SWbi}\label{APLpos1}

The object of this appendix is to calculate, for large $R$, the
boundary integrand that appears in the integral identity
\eq{weitzenbock0}, for a class of hyperboloidal initial data sets
made precise in Theorem~\ref{Tpos}. We consider a conformally
compactifiable polyhomogeneous initial data set $(\hyp,g,K)$, such
that
\bel{Kres} \mbox{$\trK$ is constant to second order at $\partial
\hyp$.}\ee
 In $\spt$ we can
always~\cite{ChMS} introduce a Bondi coordinate system $(u,x,x^A)$
such that $\hyp$ is given by an equation \bel{a1} u=\alpha(x,x^A)
\;,  \ \mbox{ with } \ \alpha(0,x^A)=0\;, \ \alpha_{,x}(0,x^A)
> 0\;,\ee where $\alpha$ is polyhomogeneous. It follows
from~\cite[Equation~(C.83)]{CJK} that \eq{Kres} is equivalent to
\bel{acond} \alpha_{,xx}|_{x=0}=0\;.\ee (Throughout this section
we will make heavy use of the formulae of~\cite[Appendix C]{CJK}
without necessarily indicating this fact.) Polyhomogeneity implies
then \bel{acond2}\alpha_{,A}=\Ol(x^3)\; ,\ee where we use the
symbol $f=\Ol(x^p)$ to denote the fact that there exists $N\in \N$
and a constant $C$ such that  $$|f|\le Cx^p(1+|\ln x|^N)\;.$$ We
also assume that this behaviour is preserved under differentiation
in the obvious way:
$$|\partial_x f|\le Cx^{p-1}(1+|\ln x|^N)\;,\quad |\partial_A f|\le Cx^p(1+|\ln x|^N)\;,$$
similarly for higher derivatives. The reason for imposing
\eq{Kres} is precisely \Eq{acond}. For more general $\alpha$'s the
expansions below acquire many further terms, and we have not
attempted to carry through the (already scary) calculations below
if \eq{acond} does not hold.

Alternatively, one can start from a space-time with a
 polyhomogeneous $\scrip$ and choose any space-like
hypersurface $\hyp$ so that \eq{a1}--\eq{acond} hold. Such an
approach can be used to study the positivity properties of the
Trautman-Bondi mass, viewed as a function on the set of
space-times rather than a function on the set of initial data
sets.

 We use the Bondi coordinates to define the background
$\fourb$: \bel{backg1} \fourb = -\rd u^2 +2x^{-2} \rd u \rd x +
x^{-2}\hst_{AB}\rd x^A \rd x^B \;,\ee
so that the components of the
inverse metric read
\[ \fourb ^{ux}=x^2 \; \, \quad \fourb ^{xx}=x^4 \; \, \quad \fourb ^{AB}=x^2\hst^{AB} \; . \]
The metric $\thb$ induced on $\hyp$ takes thus the form
\begin{eqnarray}
 \thb & = & x^{-2}\biggl[ \left(\hst_{AB} +\Ol(x^8)\right)\rd x^A \rd x^B
  +2(1-x^2\alpha_{,x})\alpha_{,A}\rd x \rd x^A \nonumber
          \\ & &  \label{3b}
+ 2\alpha_{,x}\left(1-\frac12x^2\alpha_{,x}\right)(\rd x)^2
\biggr]\;.
          \end{eqnarray}

Let ${\tilde e}^a$ be a local orthonormal co-frame for the unit
round metric $\hst$ on $S^2$ (outside of the south and north pole
one can, {\em e.g.},  use ${\tilde e}^1=\rd\theta$, ${\tilde
e}^2=\sin\theta\rd\varphi$), we set
$$e^a:=x^{-1}{\tilde e}^a\;,\quad e^3:=x^{-1}{\tilde e}^3\;,$$
where
\[ {\tilde e}^3 := \sqrt{\alpha_{,x}\left(2-x^2\alpha_{,x}\right)} \rd x
+
\frac{1-x^2\alpha_{,x}}{\sqrt{\alpha_{,x}\left(2-x^2\alpha_{,x}\right)}}
\alpha_{,A}\rd x^A \;.\] Assuming \eq{acond}-\eq{acond2}, it
follows that the co-frame $\tilde e^i$ is close to being
orthonormal for $\thb$: \be\label{3bon} \thb = x^{-2} \left[
{\tilde e}^1{\tilde e}^1 + {\tilde e}^2{\tilde e}^2+
         {\tilde e}^3{\tilde e}^3 + \Ol(x^6) \right]
         =  {e}^1{e}^1 + {e}^2{e}^2 +
         {e}^3{e}^3 + \Ol(x^4) \;. \ee
         Here and elsewhere, an equality \emph{$f=\Ol(x^p)$ for a
         tensor field $f$ means that the components of $f$ in the
         coordinates $(x,x^A)$ are $\Ol(x^p)$.}

Recall that in Bondi-Sachs coordinates \( (u,x,x^{A}) \) the
space-time metric takes the form: \be\label{gB} \fourg= -xV{\rm
e}^{2\beta}\rd u^2 + 2 {\rm e}^{2\beta} x^{-2}\rd u \rd x
   + x^{-2} h_{AB} \left(\rd x^A - U^A\rd u \right)
   \left(\rd x^B - U^B\rd u \right)\;. \ee
   This leads to the following form of the metric $\thg$ induced on
   $\hyp$
\begin{eqnarray} \hspace*{-1cm}
x^2 \thg &=& \nonumber 2\left[ (h_{AB}U^AU^B - x^3 V {\rm
e}^{2\beta})\alpha_{,x}\alpha_{,C}+
 {\rm e}^{2\beta}\alpha_{,C} - h_{CB}U^B \alpha_{,x}
\right] \rd x^C \rd x \\ \nonumber & & + \left[ (h_{AB}U^AU^B -
x^3 V {\rm e}^{2\beta})\alpha_{,D}\alpha_{,C}
   -2 h_{CB}U^B \alpha_{,D} +  h_{CD} \right] \rd x^C\rd x^D \\ & &
   + \left[ \left(h_{AB}U^AU^B-x^3V{\rm e}^{2\beta} \right)
(\alpha_{,x})^2 +2{\rm e}^{2\beta}\alpha_{,x} \right] (\rd x)^2
\label{3g}
\end{eqnarray}
Let $\gamma_{ij}$ be defined as \be\label{3gbis} x^2 \thg =
\gamma_{AB} \rd x^A \rd x^B + 2 \gamma_{xA} \rd x^A \rd x +
           \gamma_{xx} (\rd x)^2 \;,\ee
so that
\[ \gamma_{CD}:=   h_{CD} -2 \alpha_{,(D} h_{C)B}U^B
+ (h_{AB}U^AU^B - x^3 V {\rm e}^{2\beta})\alpha_{,D}\alpha_{,C}
     \;, \]
\[ \gamma_{xC}:=
(h_{AB}U^AU^B - x^3 V {\rm e}^{2\beta})\alpha_{,x}\alpha_{,C}+
 {\rm e}^{2\beta}\alpha_{,C} - h_{CB}U^B \alpha_{,x} \;,\]
\[ \gamma_{xx}:= \left[ \left(h_{AB}U^AU^B{\rm e}^{-2\beta}-x^3V \right)
\alpha_{,x} +2 \right]{\rm e}^{2\beta}\alpha_{,x} \;.\]
 If we assume that $(\hyp,g,K)$
is polyhomogeneous and conformally $C^1\times C^0$-compactifiable,
it follows that\[ h_{AB}=\hst _{AB}(1+\frac{x^{2}}{4}\chi
^{CD}\chi _{CD})+x\chi
_{AB}+x^2\zeta_{AB}+x^{3}\xi_{AB}+\Ol(x^{4})\, \, \, ,\] where
$\zeta_{AB}$ and $\xi_{AB}$ are polynomials in $\ln x$ with
coefficients which smoothly depend upon the $x^A$'s. By definition
of the Bondi coordinates we have $\det h = \det \hst$, which
implies
$$\hst^{AB}\chi_{AB}=\hst^{AB}\zeta_{AB}=0\;.$$ Further,
 \be\label{beta}
\beta= -\frac1{32}\chi^{CD}\chi_{CD}x^2 +B x^3 + \Ol(x^4) \;,\ee
 \be\label{xV} xV=1-2Mx +\Ol(x^2) \;,\ee \be\label{hU}
h_{AB}U^B = -\frac12\chi_A{^B}{_{||B}}x^2 + \newW_Ax^3
+\Ol(x^4)\;, \ee
where $B$ and $\newW_A$ are again polynomials in
$\ln x$ with smooth coefficients depending upon the $x^A$'s, while
$||$ denotes covariant differentiation with respect to the metric
$\hst$. This leads to the following approximate formulae
$$(h_{AB}U^AU^B -
x^3 V {\rm e}^{2\beta})\alpha_{,D}=\Ol(x^5)\;, \quad h_{CB}U^B
\alpha_{,D}=\Ol(x^5)\;,$$ \be\label{gxA} \gamma_{xA} =
 \alpha_{,A} + \alpha_{,x} \left[
\frac12\chi_A{^B}{_{||B}}x^2 - \newW_Ax^3
 \right] + \Ol(x^4) \;,\ee
\be\label{sgxx} \sqrt{\gamma_{xx}} = \sqrt{2\alpha_{,x}}
\left[1-\frac14(\alpha_{,x} +\frac1{8}\chi^{CD}\chi_{CD})x^2 +
\left( \frac12\alpha_{,x}M +B\right)x^3 \right] + \Ol(x^4) \;.\ee
Let $$h_{ab}=h(\tea,\teb)\;,$$ where $\tea$ is a basis of vectors
tangent to $S^2$ dual to $\tilde e^a$, and let $\mu^a{}_b$ be the
symmetric root of $h_{ab}$,
\[ \mu^a{_c} h_{ab} \mu^b{_d}=\delta_{cd}\;, \]
or, in matrix notation, \be\label{mhm} {^t\!}\mu h \mu = id \;,\ee
where ${^t\!}\mu^{c}{}_{a} = \mu^a{_c}$ stands for the transpose
of $\mu$. Let ${\tilde f}^a$, $a=1,2$, be the field of local
orthonormal co-frames for the metric $h_{AB}$ defined by the
formula \be\label{f2e} {\tilde e}^a=\mu^a{_b} {\tilde f}^b \;.\ee
We set \be\label{kof3} {\tilde f}^3:= \sqrt{\gamma_{xx}}\rd x
+\frac{\gamma_{xA}}{\sqrt{\gamma_{xx}}}\rd x^A\;,\ee so that
\be\label{3gon} x^{2} \thg  =  {\tilde f}^1{\tilde f}^1 + {\tilde
f}^2{\tilde f}^2+
         {\tilde f}^3{\tilde f}^3 + \Ol(x^4)  \;.\ee
There exists a matrix $M^k{_l}$ such that
\[ {\tilde e}^k = M^k{_l}{\tilde f}^l\;.\]
Now, \be\label{f3e} {\tilde e}^3= (1-x^2 a){\tilde f}^3 - x^2 b_a
\mu^a{_b} {\tilde f}^b
 +\Ol(x^4)\;,
\ee with
\[ a:= - \frac1{32}\chi^{CD}\chi_{CD}
+ \left( \frac12\alpha_{,x}M + B\right)x \;,\]
\[ b_a \tilde e^a = b_A dx^A:=\sqrt{\frac{ \alpha_{,x}}2} \left[
\frac12\chi_A{^B}{_{||B}} - \newW_Ax
 \right] dx^A\;. \]
 The matrix $M$ is easily calculated to be
\be\label{M33} M^3{_3} = 1-x^2 a +\Ol(x^4)\; , \quad M^3{_b} = -
x^2 b_a \mu^a{_b} +\Ol(x^4)\;, \ee \be\label{Mab} M^a{_3} = 0 \; ,
\quad M^a{_b} = \mu^a{_b}\;.\ee Since
\[ e^k = x^{-1} {\tilde e}^k \; , \quad f^k = x^{-1} {\tilde f}^k \;,\]
it follows that \[ e^k = M^k{_l} f^l \; , \quad f_l = M^k{_l}
e_k\;,
\] where $f_l$ and  $e_k$  stand for frames dual to $f^i$ and $e^i$. If
we choose
\[ {\tilde e}^1=\rd\theta \; , \quad
{\tilde e}^2=\sin\theta\rd\varphi \;,\] then
\[ {\tilde e}^3 := \sqrt{\alpha_{,x}\left(2-x^2\alpha_{,x}\right)} \rd x
+
\frac{1-x^2\alpha_{,x}}{\sqrt{\alpha_{,x}\left(2-x^2\alpha_{,x}\right)}}
\alpha_{,A}\rd x^A \;,\]
\[ {\tilde e}_1=\frac{\partial}{\partial\theta}
   -\alpha_{,\theta}
\frac{1-x^2\alpha_{,x}}{\alpha_{,x}\left(2-x^2\alpha_{,x}\right)}
   \frac{\partial}{\partial x} \;,\]
\[ {\tilde e}_2=\frac1{\sin\theta}\left[\frac{\partial}{\partial\varphi}
   -\alpha_{,\varphi}
\frac{1-x^2\alpha_{,x}}{\alpha_{,x}\left(2-x^2\alpha_{,x}\right)}
   \frac{\partial}{\partial x} \right] \;,\]
\[ {\tilde e}_3=
\frac{1}{\sqrt{\alpha_{,x}\left(2-x^2\alpha_{,x}\right)}}
   \frac{\partial}{\partial x}  \;,\]
\[ e_k = x {\tilde e}_k \; , \quad f_k = x {\tilde f}_k \;. \]
We also have the relation
\[ e_i = (M^{-1})^k{_i} f_k \;,\]
with \be\label{Mo3k} (M^{-1})^3{_3} =\frac1{M^3{_3}} = 1+x^2 a +
\Ol(x^4)\; , \quad (M^{-1})^a{_b} = (\mu^{-1})^a{_b}\;, \ee
\be\label{Moak} (M^{-1})^a{_3} = 0 \; , \quad (M^{-1})^3{_b} = -{
M^3{_a}(\mu^{-1})^a{_b}\over M^3{_3}}=
  x^2 b_b + \Ol(x^4)\;.
\ee
\newcommand{\lapse}{W}
Consider, now, the integrand in \eq{weitzenbock0}:
\[
B^3= -< \psi, \nabla^3 \psi+ \gamma^3\gamma^i \nabla_i\psi> =-
   < \psi, \gamma^3\gamma^a \nabla_a\psi>
\; , \] where the minus sign arises from the fact that we will use
a $g$--orthonormal  frame $\repf_i$ in which $\repf_3$ is
\emph{minus} the outer-directed normal to the boundary. Here
$\psi$ is assumed to be the restriction to $\hyp$ of a space-time
covariantly constant spinor with respect to the background metric
$\fourb$:
$$\znabla \psi = 0\;.$$ It follows that
\begin{eqnarray} 
\nabla_a\psi & = & \repf_a(\psi) +
\left[ -\frac12 K(\repf_a,\repf_j)\gamma^j\gamma_0 -\frac14
\omega_{ij}(\repf_a)\gamma^i\gamma^j \right] \psi  \\ & = & \left[
-\frac12 \left( K(\repf_a,\repf_j) - \zK(\repf_a,\repe_j)
\right)\gamma^j\gamma_0 + \frac14 \left( \zomega_{ij}(\repf_a) -
\omega_{ij}(\repf_a) \right) \gamma^i\gamma^j \right]\psi\nonumber
\; .
\end{eqnarray}
This allows us to rewrite $B^3$ as
\begin{eqnarray}
B^3 & = & \frac12 \left( K(\repf_a,\repf_j) - \zK(\repf_a,\repe_j)
\right) < \psi, \gamma^3\gamma^a \gamma^j\gamma_0\psi> \nonumber \\
& & - \nonumber \frac14 \left( \zomega_{ij}(\repf_a) -
\omega_{ij}(\repf_a) \right)
   < \psi, \gamma^3\gamma^a \gamma^i\gamma^j \psi> \\ &=&
 \frac12 \left( K(\repf_a,\repf_j) - \zK(\repf_a,\repe_j) \right)
 \left( g^{j3}Y^a -g^{ja} Y^3 \right) \nonumber\\ & &
+  \frac12 \left( \zomega_{3a}(\repf_a) - \omega_{3a}(\repf_a)
\right) \lapse  \; ,\label{bterms}\end{eqnarray} where
$(\lapse,Y^i)$ denotes the KID\footnote{To avoid a clash of
notation with the Bondi function $V$ we are using the symbol $W$
for the normal component of the KID here.} associated to the
spinor field $\psi$
\[ \lapse := <\psi , \psi>\;, \quad Y^k = <\psi, \gamma^k\gamma_0\psi > \; . \]
We use the convention in which
\[ \{ \gamma^i, \gamma^j \} = -2 \delta^{ij} \; , \]
with $\gamma^0$ --- anti-hermitian, and $\gamma^i$ --- hermitian.
Because $\psi$ is covariantly constant, $\lapse$ and $Y^i$ satisfy
the following equation
$$\partial_i \lapse = K_{ij}Y^j\;.$$ Let $\repe_i$ be
an orthonormal frame for $b$; we will shortly see that we have the
following asymptotic behaviors,
\[ \zomega_{3a}(\repf_a) - \omega_{3a}(\repf_a) =O (x^2) \; ,
\]
\[ K(\repf_a,\repf_j) - \zK(\repf_a,\repe_j) =O (x^2) \; , \]
\[ Y^k=O (x^{-1}) \; , \quad \lapse =O (x^{-1}) \; , \]
which determines the order to which various objects above have to
be expanded when calculating $B^3$. In particular, these equations
show that some non-obvious cancelations have to occur for the
integral of $B^3$ to converge.

Since $\repe_i$ is $b$--orthonormal, it holds that
\be\label{oko} 2\zomega_{klj} =
b([\repe_k,\repe_l],\repe_j)+b([\repe_k,\repe_j],\repe_l) -
b([\repe_l,\repe_j],\repe_k) \; . \ee It follows from \eq{3bon}
that we can choose $\repe_k$ so that \bel{reper} \repe_k =
(\delta_k^\ell + \Ol(x^6))e_\ell\;. \ee This choice leads to the
following commutators:
 \be\label{e12}
[\repe_1,\repe_2]= -x\cot\theta \repe_2 +\Ol (x^3)\repe_1 + \Ol
(x^3)\repe_2+\Ol (x^4)\repe_3 \; , \ee
\begin{eqnarray}\nonumber [\repe_a,\repe_3]&=& -\left[\alpha_{,x}(2-x^2
\alpha_{,x}) \right]^{-1/2} \repe_a
  -(\frac{\alpha_{,a}}{2\alpha_{,x}}+x\frac{\alpha_{,xa}}{4\alpha_x}+\Ol (x^4) )\repe_3 \\
  &&+ \Ol (x^4)\repe_1 + \Ol (x^4)\repe_2
\; , \label{e12.1}\end{eqnarray}
 \be\label{cijk} [\repe_i,\repe_j]
= c_{ij}{^k}\repe_k \; , \ee
\begin{eqnarray}\label{c122x}&
c_{21}{^1}= - c_{12}{^1} = \Ol(x^3)\;, &\\\label{c122}&
c_{21}{^2}= - c_{12}{^2} = x\cot\theta +\Ol(x^3)\;, &\\
\label{Ca22}
& c_{12}{}^3=-c_{21}{}^3= \Ol(x^4)\;,& \\
 \label{ca33}&
c_{3a}{^3}= - c_{a3}{^3} =  \frac{\alpha_{,a}}{2\alpha_{,x}}+x\frac{\alpha_{,xa}}{4\alpha_x} +\Ol(x^4)\;,&\\
\label{ca3b} &c_{3a}{^b}= - c_{a3}{^b} =  \left[\alpha_{,x}(2-x^2
\alpha_{,x}) \right]^{-1/2}\delta^b{_a} +\Ol(x^4)  \; .&
\end{eqnarray}
 It follows from
\eq{3gon} that we can choose $\repf_j$ so that \bel{repfr} \repf_j
= (\delta_j^k + \Ol(x^4))f_k\;.\ee
 Let $\hat M$ be the
transition matrix from the frame $\repe_i$ to the frame $\repf_j$,
$$ \repf_k = \hat M ^\ell {}_k \repe_\ell\; ,$$ if we define the functions $d_{ij}{}^k$ as
\be\label{dijkf} [\repf_i,\repf_j] = d_{ij}{^k}\repf_k \; , \ee
then we have \be\label{dlmk}
 d_{ml}{^k}= {\hat M}^j{_l} {\hat M}^i{_m}c_{ij}{^n}({\hat M}^{-1})^k{_n} +
 ({\hat M}^{-1})^k{_j}\repf_m ({\hat M}^j{_l})
 - ({\hat M}^{-1})^k{_j}\repf_l ({\hat M}^j{_m})
\; . \ee Chasing through the definitions one also finds that
\bel{Om3afa0}\hat M^i{}_j = M^i{}_j + \Ol(x^4)\;.\ee
This leads to
$$d_{3aa} =  M^3{}_3c_{3aa}+\Ol (x^4) =2\left[\alpha_{,x}(2-x^2
\alpha_{,x}) \right]^{-1/2} M^3{}_3+\Ol (x^4)\;.$$
Now, \be\label{ockl}
2\zomega_{klj} 
= c_{klj} + c_{kjl} - c_{ljk} \; , \ee \be\label{odkl}
2\omega_{klj} 
= d_{klj} + d_{kjl} - d_{ljk} \; , \ee \be\label{m3ab}
\omega_{3ab} = d_{3(ab)}-\frac12 d_{ab3} \; , \ee \be\label{m3aa}
\omega_{3aa} = d_{3aa} \; , \ee \be\label{om3ab} \zomega_{3ab} =
\left[\alpha_{,x}(2-x^2 \alpha_{,x}) \right]^{-1/2}\delta_{ab}+\Ol
(x^4) = O(1) \; , \ee \be\label{om3a3} \zomega_{3a3} =
\Ol (x^3) \; . \ee Using obvious matrix notation, from \eq{mhm} we
obtain \bel{mudef} \mu = \id -\frac x 2 \chi +x^2 d + x^3 w +
\Ol(x^4)\;,\ee  where $d$ and $w$ are polynomials in $\ln x$ with
coefficients which are $x^A$--dependent symmetric matrices. The
condition that $\det \mu= 1$ leads to the relations
$$\tr \chi = 0\;,\quad  \tr d = \frac 18 \tr \chi^2\;, \quad \tr w
= - \frac 12 \tr(\chi d)\;. $$ Using $(\mu^{-1})^a{}_b
\repf_3(\mu^b{}_a)=\repf_3(\det \mu )= 0$, the asymptotic
expansions \eq{c122}-\eq{ca3b} together with \eq{Mo3k}-\eq{Moak},
\eq{reper} and \eq{repfr} one finds the following contribution to
the first term $\frac12 \left( \zomega_{3a}(\repf_a) -
\omega_{3a}(\repf_a) \right) \lapse $ in $B^3$:
\begin{eqnarray}\label{om3afa}
\zomega_{3a}(\repf_a) - \omega_{3a}(\repf_a)& = & \hat
M^l{_a}\zomega_{3al} - \omega_{3aa}\nonumber \\ & = &
M^3{_a}\zomega_{3a3} + (\mu^b{_a}-\delta^b{_a})\zomega_{3ab} +
\zomega_{3aa} - \omega_{3aa} + \Ol(x^4)\nonumber
\\ 
& =&
 \frac{x^2}{\sqrt{2\alpha_{,x}}} \left[
\underline{\frac1{16}\chi_{AB}\chi^{AB}}
 +x\left( M \alpha_{,x} +2B -\frac 12 \tr (\chi d)
 \right) \right]\nonumber\\&&\mbox{}+ \Ol(x^4)
\; .\end{eqnarray}
 The underlined term would give
a diverging contribution to the integral of $B^3$ over the
conformal boundary if it did not cancel out with an identical term
from the $K$ contribution, except when $\chi=0$.

We choose now the Killing spinor $\psi$ so that $$\lapse \zn + Y =
\frac{\partial}{\partial t}$$ in the usual Minkowskian
coordinates, where $\zn$ is the unit normal to $\hyp$. This leads
to
\[ \lapse =\frac{x^{-1}}{\sqrt{2\alpha_{,x}}}\left[ 1+O (x^2)\right]
\; , \quad Y^x = \frac{1}{2\alpha_{,x}} \left[ 1+O (x^2)\right]
 \; , \quad Y^A = O (x^2) \; , \]
\[ Y^3 = \frac{x^{-1}}{\sqrt{2\alpha_{,x}}} +O (x)
 \; , \quad Y^a = O (x) \; . \]
 In order to calculate the remaining terms in
 \eq{bterms} we begin with
\begin{equation}
 K(\repf_a,\repf_a) - \zK(\repf_a,\repe_a)  =
x^2\mu^c{_a} \mu^d{_a} K(\tilde e_c,\tilde e_d)
  - x^2\mu^c{_a}\zK(\tilde e_c,\tilde e_a)  +\Ol (x^4)
\; , \end{equation} Using the formulae of \cite[Appendix C.3]{CJK}
one finds \[ \mu^c{_a} \mu^d{_a} K(\tilde e_c,\tilde
e_d)=h^{cd}K(\tilde e_c,\tilde e_d)=h^{AB}K_{AB} + \Ol (x^2) \; ,
\] \bel{gamome1} \Gamma^\omega{_{AB}}=
   \Gamma^u{_{AB}} -\alpha_{,x} \Gamma^x{_{AB}} + \Ol (x^3) \; , \ee
\be \Gamma^u{_{AB}} = x^{-1}{\rm e}^{-2\beta}
    \left( h_{AB} -\frac x2 h_{AB,x}\right) \; , \ee
\be \Gamma^x{_{AB}} = -\frac12{\rm e}^{-2\beta}
    \left( 2  \mcD_{(A}U_{B)} +\partial_u h_{AB}
    -2Vx^2h_{AB} + Vx^3 h_{AB,x}\right) \; , \ee
where $ \mcD_{A}$ denotes the covariant derivative with respect to
the metric $h$. It follows that \bel{gamome}
\Gamma^\omega{_{AB}}h^{AB}=
    x^{-1}{\rm e}^{-2\beta}
    \left[ 2+ x\alpha_{,x}\left( U^A{}_{||A} -2Vx^2\right)
    \right] +\Ol (x^3)
\; . \ee Further
\[ N= {x^{-1}\over \sqrt{2\alpha_{,x}}} \left[
   1+\beta +\frac14 x^3 \alpha_{,x}V + O(x^4) \right] \; . \]
   \Eq{gamome1} specialised to Minkowski metric reads\be
 {\mathring \Gamma}^\omega{_{AB}}= x^{-1}\hst_{AB} -
  x\alpha_{,x}\hst_{AB} +\Ol(x^3)
\; , \ee yielding
\begin{eqnarray}
\zK_{AB} & = & -\frac{x^{-2}}{\sqrt{2\alpha_{,x}}} \biggl[
\hst_{AB}\left( 1 - \frac34 x^2\alpha_{,x} \right)
  + \Ol (x^4) \biggr] \label{KoAB}
 \; .\end{eqnarray}
We therefore have\begin{eqnarray} x^2h^{AB}K_{AB}
-x^2\mu^b{_a}\zK_{ab} & = &   \frac1{\sqrt{2\alpha_{,x}}} \Big[
   2\beta+\frac18x^2\chi_{CD}\chi^{CD} -\frac32x^2\alpha_{,x}(1-xV)\nonumber \\ & &
   \mbox{}+x\alpha_{,x}U^A{_{||A}}-\frac {x^3}2 \tr(\chi d) \Big] + \Ol (x^4) \; , \end{eqnarray}
   which implies
\begin{eqnarray} 
\left( K(\repf_a,\repf_a) - \zK(\repf_a,\repe_a)\right)Y^3 & =&
\nonumber \frac{x}{2\alpha_{,x}} \biggl[ \underline{
\frac1{16}\chi_{AB}\chi^{AB} } + \\ & & \hspace*{-3.5cm}
 \mbox{}+ x \left( 2B -3 M\alpha_{,x} + \frac12\alpha_{,x}\chi^{AB}{_{||AB}}
 - \frac 12 \tr(\chi d)
  \right)  \biggr] + \Ol (x^3)\;.
  \nonumber \\ &&
 \label{B3KY}
\end{eqnarray}
Here we have again underlined a potentially divergent term. Using
the formulae of \cite[Appendix~C]{CJK} one further finds
 \[  \left( K(\repf_a,\repf_3) - \zK(\repf_a,\repe_3) \right) Y^a  = \Ol (x^3) \; , \]
which will give a vanishing contribution to $B^3$ in the limit.
 Collecting
this together with \eq{B3KY} and \eq{om3afa} we finally obtain
\begin{equation}
 B^3
   =  \frac14 x^2 \left( 4M -  \frac12\chi^{AB}{_{||AB}} \right)
   +\Ol(x^3)
\; , \end{equation} We have thus shown that the integral of $B^3$
over the conformal boundary is proportional to the Trautman-Bondi
mass, as desired.

\section{Proof of Lemma~\ref{Lpos2}}
\label{APLpos2} From
$$\znabla \psi = 0$$ we have
\begin{eqnarray} 
\gamma^\ell\nabla_\ell\psi & = & \gamma^\ell\repf_\ell(\psi) +
\gamma^\ell\left[ -\frac12 K(\repf_\ell,\repf_j)\gamma^j\gamma_0
-\frac14
\omega_{ij}(\repf_\ell )\gamma^i\gamma^j \right] \psi \nonumber \\
& = & \left[ -\frac12 \left( K(\repf_\ell ,\repf_j) -
\zK(\repf_\ell ,\repe_j) \right)\gamma^\ell\gamma^j\gamma_0 \right.
\nonumber \\
&& \left. +
\frac14 \left( \zomega_{ij}(\repf_\ell ) - \omega_{ij}(\repf_\ell
) \right) \gamma^\ell\gamma^i\gamma^j \right]\psi \; .\label{c1}
\end{eqnarray}
We start by showing that
\[  \left[ \zomega_{ij}(\repf_l)-\omega_{ij}(\repf_l) \right]\gamma^l
\gamma^i\gamma^j =
     O(x^2) \;.\]
     In order to do that, note first
\[ \gamma^i\gamma^j\gamma^k \Delta_{jki} = \varepsilon^{ijk}\Delta_{jki}\gamma^1\gamma^2
\gamma^3 - 2g^{ij}\gamma^k\Delta_{jki} \;,\] where
\[
\Delta_{jki}:=\zomega_{jk}(\repf_i)-\omega_{jk}(\repf_i)=
 {\hat M}^l{_i}\zomega_{jkl} - \omega_{jki}\;.\]We claim that
\bel{cclaim} \varepsilon^{ijk}\Delta_{jki}=O(x^2)\, , \quad
   \Delta_{jkk}=O(x^2)\;.\ee
   The intermediate calculations needed for this are as follows:
\be\label{e12x2}
[\repe_1,\repe_2]= -x\cot\theta \repe_2 +\Ol (x^4) \ee (which
follows from \eq{e12}),
\be\label{ea3x2}
[\repe_a,\repe_3]= -\left[2\alpha_{,x} \right]^{-1/2} \repe_a
   +\Ol (x^4)\;,
\ee
\be\label{f12x2}
[\repf_1,\repf_2]= -x\cot\theta \repf_2 +O(x^2)\;, \ee
\be\label{fa3x2}
[\repf_3,\repf_a]= \left[2\alpha_{,x} \right]^{-1/2} \left(
\repf_a -\frac12 x\chi^b{_a}\repf_b \right)
   +O(x^2)\;.
\ee In order to calculate $ \Delta_{123} + \Delta_{312} +
\Delta_{231}$, we note that \beaa \lefteqn{\zomega_{12}(\repf_{3})
+ \zomega_{31}(\repf_{2}) + \zomega_{23}(\repf_{1}) =
\zomega_{123}
+ \mu_2{^a} \zomega_{31a} + \mu_1{^a} \zomega_{23a} + O(x^2)}& & \\
&&=
 (\zomega_{123} +  \zomega_{312} +  \zomega_{231})
 -\frac12 x \left( \chi_2{^a}\zomega_{31a}+ \chi_1{^a}\zomega_{23a} \right)
  + O(x^2)\;.\eeaa
From
  $\zomega_{3ab}=\left[2\alpha_{,x} \right]^{-1/2}\delta_{ab} + \Ol(x^3)$ the
  $\chi$ terms drops out. Equations~(\ref{e12x2})-(\ref{ea3x2})
  show that $$2\zomega_{123} +  2\zomega_{312} +  2\zomega_{231} =
 b([\repe_2,\repe_1],\repe_3)+ b([\repe_3,\repe_2],\repe_1) +
 b([\repe_1,\repe_3],\repe_2)= O(x^2)\;.$$
Similarly it follows from  (\ref{f12x2})-(\ref{fa3x2}) that
$$2\omega_{123} +  2\omega_{312} +  2\omega_{231} =
 g([\repf_2,\repf_1],\repf_3)+ g([\repf_3,\repf_2],\repf_1) +
 g([\repf_1,\repf_3],\repf_2)= O(x^2)\;,$$
yielding finally $$ \Delta_{123} + \Delta_{312} + \Delta_{231} =
O(x^2)\;.$$ Next, $\Delta_{3kk}=\Delta_{3aa}$ is given by
(\ref{om3afa}), and is thus $O(x^2)$. We continue with
$\Delta_{akk}=\Delta_{abb}+ \Delta_{a33}$. For $a=1$ one finds
$$\omega_{122} +\omega_{133}=g([\repf_1,\repf_2],\repf_2) +
g([\repf_1,\repf_3],\repf_3)= -x\cot\theta +O(x^2)\;.$$ Further,
$$\zomega_{11}(\repf_1)+\zomega_{12}(\repf_2)+\zomega_{13}(\repf_3)=
  \zomega_{133}+\mu_a{^b}\zomega_{1ab} +O(x^2) =
  \zomega_{1kk} -\frac12 x \chi_a{^b} \zomega_{1ab} +O(x^2)\;.$$
 We have $\zomega_{122}= -x\cot\theta + \Ol(x^3)= -x\cot\theta + O(x^2)$,
 while
  $\zomega_{1ab}=O(x^2)$ as well, so that the $\chi$'s can be absorbed in the error terms, leading to
$$\Delta_{1kk}=
\zomega_{11}(\repf_1)+\zomega_{12}(\repf_2)+\zomega_{13}(\repf_3)
- ( \omega_{122} +\omega_{133}) =O(x^2)\;.$$ For $a=2$ we
calculate
$$\omega_{211} +\omega_{233}=g([\repf_2,\repf_1],\repf_1)
+ g([\repf_2,\repf_3],\repf_3)= O(x^2)\;,$$ and it is easy to
check now that $\Delta_{2kk}=O(x^2)$. This establishes
\eq{cclaim}.

To estimate the contribution to \eq{c1} of the terms involving $K$
we will need the following expansions
\be\label{d3abx2}
d_{3a}{^b} = \frac{1}{\sqrt{2\alpha_{,x}}} \left[ \delta^b{_a}
-\frac12 x\chi^b{_a}  \right] +\Ol(x^2)\;, \ee
\be\label{d3a3x2}
d_{3a}{^3} = -\frac{x^2b_a}{\sqrt{2\alpha_{,x}}} + \Ol(x^3) =
\Ol(x^2)\;, \ee
\be\label{dab3x2}
d_{ab}{^3} = x^2 c_{ab}{^c}b_c  +\Ol(x^3)= O(x^2)\;. \ee Now, it
follows from \cite[Appendix~C.3]{CJK} that \[ tr_g K -
\zK(\repf_i,\repe_i) = \Ol(x^2)\;. \] Next, we claim that
\[ \zK(\repf_j,\repe_k)- \zK(\repf_k,\repe_j)=\Ol(x^2) \;.\]
We have the following asymptotic formulae for the connection
coefficients:
\[ \Gamma^\omega{_{xx}} = 2x^{-1}\alpha_{,x} + O(x)\;, \quad  \Gamma^\omega{_{xA}} = O(x)\;,\]
\[
\Gamma^\omega{_{AB}} = \nonumber x^{-1}\hst_{AB} +\frac12\chi_{AB}
+ O(x) \;,\] and for the extrinsic curvature tensor:
\[ K_{xx} = -N \Gamma^\omega{_{xx}}= - x^{-2} \left[
\sqrt{2\alpha_{,x}} +O(x^2) \right] = \zK_{xx}\;, \]
\[ K_{xA} = -N \Gamma^\omega{_{xA}}= O(1)  \; , \quad \zK_{xA}=O(x) \;,\]
\begin{eqnarray} \nonumber
K_{AB} = -N \Gamma^\omega{_{AB}} & = &
-\frac{x^{-2}}{\sqrt{2\alpha_{,x}}} \biggl[
\hst_{AB}+\frac12x\chi_{AB} + O(x^2) \biggr] \;,\label{KABx2}
 \end{eqnarray}
\begin{eqnarray}
\zK_{AB} & = & -\frac{x^{-2}}{\sqrt{2\alpha_{,x}}} \biggl[
\hst_{AB}
  + O(x^2) \biggr] \;.\label{KoABx2}
 \end{eqnarray}
  From the above and~\cite[Appendix~C.3]{CJK} one
obtains the following formulae
\be \zK(\repe_k,\repe_l) = -\frac{1}{\sqrt{2\alpha_{,x}}}
\delta_{kl}  +\Ol (x^2) \; , \ee
\be  K(\repe_k,\repe_l) = -\frac{1}{\sqrt{2\alpha_{,x}}}
\biggl[ \delta_{kl} +\frac12x\chi_{kl}
  \biggr]  +\Ol (x^2) \; , \ee where we have set
  $\chi_{3k}=0$. Further
\[ {\hat M}^l{_k} = \delta^l{_k} -\frac12 x \chi^l{_k}+\Ol (x^2)\;, \]
\beaa \lefteqn{\zK(\repf_j,\repe_k)- \zK(\repf_k,\repe_j) =
   {\hat M}^l{_j}\zK(\repe_l,\repe_k)-
   {\hat M}^l{_k}\zK(\repe_l,\repe_j)}&&\\ & & =
   -\frac{1}{\sqrt{2\alpha_{,x}}}
   \left(\delta^l{_j}-\frac12 x \chi^l{_j}\right)
  \delta_{lk}
  +\frac{1}{\sqrt{2\alpha_{,x}}}
  \left(\delta^l{_k}-\frac12 x \chi^l{_k}\right)
  \delta_{lj}
  + \Ol (x^2)\\ &&= \Ol (x^2)\;.
\eeaa This, together with \cite[Equation~(C.83)]{CJK} yields
$$ tr_g K - \zK(\repf_i,\repe_i)= -\frac{3}{\sqrt{2\alpha_{,x}}}
+ \frac{1}{\sqrt{2\alpha_{,x}}} \left(\delta^l{_i}-\frac12 x
\chi^l{_i}\right)
  \delta_{li}
+\Ol(x^2) = \Ol(x^2)\;.$$ We also have
\begin{eqnarray} \nonumber 
\left[ K(\repf_l,\repf_j) - \zK(\repf_l,\repe_j) \right]
\gamma^l\gamma^j
 & = &
 -K(\repf_k,\repf_k) - \zK(\repf_l,\repe_j)\gamma^l \gamma^j
  \\
 & = & -tr_g K + \zK(\repf_k,\repe_k)  \nonumber\\ &&
+\frac12\left(\zK(\repf_j,\repe_k)-
\zK(\repf_k,\repe_j)\right)\gamma^k \gamma^j\;, \nonumber
\end{eqnarray} so that
\[ \left[ K(\repf_l,\repf_j) - \zK(\repf_l,\repe_j) \right]
\gamma^l\gamma^j= \Ol(x^2)\;.
\]
Since
\[ \sqrt{\det g}=O(x^3) \, , \quad <\psi,\psi>=O(x^{-1})\;,\]
we obtain\[ \sqrt{\det g}\left| \gamma^k \nabla_k\psi\right|^2
=\Ol(1)\in L^1\;.
\]

\section{Asymptotic expansions of objects on $\hyp$}

Throughout this appendix coordinate indices are used.
\subsection{Smooth case}

\label{app:obiekty3d}
\paragraph{Induced metric \protect\( \gmetric \protect \).
} We write the spacetime metric in Bondi-Sachs coordinates, as in
\eq{gB},  and
 use the standard expansions for the coefficients of the metric
 (see \emph{e.g.}\/~\cite[Equations~(5.98)-(5.101)]{CJK}). Let \( \hyp  \) be given by \( \omega
=\mathrm{const}., \) where
\[ \omega =u-\alpha (x,x^{A})\;.\]
Two different coordinate systems will be used: \( (u,x,x^{A}) \)
and \( (\omega ,x,x^{A}) \) --- coordinates adapted to \( \hyp .
\) To avoid ambiguity two different symbols for partial
derivatives will be used: the comma stands for the derivative with
\( \omega=\mathrm{const}. \) and \( \uder  \) stands for the
derivative with \( u=\mathrm{const}. \) These two derivatives can
be transformed into each other:\[ A_{,x}=A_{\uder x}+\alpha
_{\uder x}\partial _{u}A\;,\]
\[
A_{,A}=A_{\uder A}+\alpha _{\uder A}\partial _{u}A\;.\] For
functions not depending on \( u \) ({\em e.g.,}\/ \( \alpha  \))
the symbols mean the same. Derivations in covariant derivatives \(
_{||A} \) and \( _{|i} \) are with \( \omega=\mathrm{const}.  \)

\paragraph{Three-dimensional reciprocal metric.
} We have the following implicit formulae for the three
dimensional inverse metric \( \gmetric ^{ij} \):\[ -\frac{\gmetric
^{xA}}{\gmetric ^{xx}}=\gtw ^{AB}\gmetric _{xB}\;,\]
\[
\frac{1}{\gmetric ^{xx}}=\gmetric _{xx}+\frac{\gmetric
^{xA}}{\gmetric ^{xx}}\gmetric _{xA}\;,\]
\[
\gmetric ^{AB}=\gtw ^{AB}+\frac{\gmetric ^{xA}\gmetric
^{xB}}{\gmetric ^{xx}}\;,\] where $\gtw^{AB}$ denotes the matrix
inverse to $(g_{AB})$. The calculations get very complicated in
general. To simplify them we will assume a particular form of \(
\alpha , \) \emph{i.e.},
\begin{equation} \label{alfa} \alpha =\mathrm{const}.\cdot
x+O(x^{3})\;.
\end{equation}
This choice is motivated by the form of \( \alpha  \) for standard
hyperboloid \( t^{2}-r^{2}=1 \) in Minkowski spacetime. In that
case \( \alpha
=\frac{1}{x}\sqrt{1+x^{2}}-\frac{1}{x}=\frac{1}{2}x+O(x^{3}). \)
It is further equivalent to the asymptotically CMC condition
\eq{fr1}. With the above assumptions we have the following
asymptotic expansions:
\[ \gtw^{AB}=x^{2}h^{AB}+O(x^7)\;,\]
 \begin{eqnarray} \nonumber
-\frac{\gmetric ^{xA}}{\gmetric ^{xx}}&=& \alpha ^{,A}+x^{2}\alpha
_{,x} \left(\frac{\chi^{AC}{_{||C}}}{2}-2N^{A}x-\frac{(\chi
^{CD}\chi _{CD})^{||A}}{16}x -\frac{\chi
^{AB}\chi_{B}{^C}{_{||C}}}{2}x\right)\\ && +O(x^{4})\;,
\nonumber \end{eqnarray}
\[
\frac{1}{\gmetric ^{xx}}=\frac{2\alpha _{,x}}{x^{2}}\left[1+2\beta
-\frac{x^{2}\alpha _{,x}}{2}+Mx^{3}\alpha
_{,x}+O(x^{4})\right]=\gmetric _{xx}\;,\]
\[
\gmetric ^{xx}=\frac{x^{2}}{2\alpha _{,x}}\left[1-2\beta
+\frac{x^{2}\alpha _{,x}}{2}-Mx^{3}\alpha
_{,x}+O(x^{4})\right]\;,\]
\[
\frac{1}{\sqrt{\gmetric ^{xx}}}=\frac{\sqrt{2\alpha _{,x}}}{x}
\left[1+\beta -\frac{1}{4}x^{2}\alpha
_{,x}(1-2Mx)+O(x^{4})\right]\;,\]
\[
-\gmetric ^{Ax}=\frac{x^{4}}{2}\left[\frac{\chi^{AC}{_{||C}}}{4}
-2xN^{A}-x\frac{(\chi ^{CD}\chi _{CD})^{||A}}{16} -x\frac{\chi
_{C}{^A}\chi^{CD}{_{||D}}}{2} +\frac{\alpha _{,A}}{x^{2}\alpha
_{,x}}+O(x^{2})\right]\;,\]
\[
g ^{AB}=x^{2}h^{AB}+O(x^{6})\;.\]

\paragraph{Determinant of the induced metric $\gtw$.}
 The determinant of the 2-dimensional metric induced on
the Bondi spheres $u=\const.$, $x=\const.$, equals
\[
\det \big(\frac{1}{x^2} h\big) = \det \big(\frac{1}{x^2}
\hst\big)\;.
\]
In our case the $\omega=\const.$, $x=\const.$ spheres are not
exactly the Bondi ones, and one finds
\[
\lambda :=\sqrt{\det \gtw } = \frac{\sqrt{\hst}}{x^2} + O(x^3)\;,
\quad \zlambda :=\sqrt{\det \btw } = \frac{\sqrt{\hst}}{x^2} +
O(x^6)\;.
\]
We also note the following purely algebraic identities:
\begin{equation} \label{lgg} \lambda =\sqrt{\gmetric
^{xx}}\cdot \sqrt{\det \gmetric}\;, \quad \zlambda
=\sqrt{b^{xx}}\cdot \sqrt{\det b}\;,
\end{equation}

\paragraph{Lapse and shift.} The function \( N \) (the \emph{lapse}) and the vector \( S_{i}
\) (the \emph{shift}) can be calculated as follows:\[
N=\frac{1}{\sqrt{-\fourg^{\omega \omega }}}\;,\]
 \[
S_{i}=\fourg_{\omega i}\;.\] Hence \[ N=\frac{1}{x\sqrt{2\alpha
_{,x}}}\left[1+\beta +\frac{1}{4}x^{2}\alpha _{,x}
-\frac{1}{2}x^{3}\alpha _{,x}M+O(x^{4})\right]\;,\]
 \[
S_{x}=\frac{1}{x^{2}}\left[1+2\beta -x^{2}\alpha
_{,x}+2x^{3}\alpha _{,x}M+O(x^{4})\right]\;,\]
 \[
S_{A}=\frac{1}{2}\chi^{C}{_{A||C}}-2xN_{A}-\frac{1}{16}x(\chi
^{CD}\chi _{CD})_{||A}+O(x^{2})\;.\]

\paragraph{Christoffel coeficients.
}

Using the induced metric (and the reciprocal metric) we can
compute the coefficients \(\Gtrzy ^{i}{_{jk}}\). The results are:

\label{Gtrzy}

\[
\Gtrzy ^{x}{_{xx}}=-\frac{1}{x}+\frac{\alpha _{,xx}}{2\alpha
_{,x}}+\beta _{\uder x} +\alpha
_{,x}\left(\frac{3}{2}x^{2}M-\frac{1}{2}x+\partial _{u}\beta
\right)+O(x^{3})\;,\]
\begin{eqnarray*}
\Gtrzy ^{x}{_{xA}} & = &
\frac{1}{4}x\left[\chi^{C}{_{A||C}}-4xN_{A}
-\frac{1}{8}x(\chi ^{CD}\chi _{CD})_{||A}-\frac{1}{2}x\chi _{AC}\chi^{CD}{_{||D}}
\right.\\
 &  & \left.{}+\frac{2\alpha _{,A}}{x^{2}\alpha _{,x}}+\frac{4\beta _{\uder A}}{x}
 +\frac{2\alpha _{,xA}}{x\alpha _{,x}}+O(x^{2})\right]\;,\nonumber
\end{eqnarray*}
\[
\Gtrzy ^{x}{_{AB}}=\frac{\hst _{AB}}{2x\alpha _{,x}} +\frac{\chi
_{AB}}{4\alpha _{,x}}+\frac{1}{4}x\hst _{AB}-\frac{1}{4}x\partial
_{u}\chi _{AB} -\frac{\beta \hst _{AB}}{x\alpha _{,x}}+O(x^{2})\;,\]
\[
\Gtrzy ^{A}{_{xx}}=\frac{1}{2}x\alpha _{,x}\chi
^{AC}{_{||C}}+O(x^{2})\;,\]
\begin{eqnarray*}
\Gtrzy ^{A}{_{xB}} & = &
-\frac{1}{x}\delta^{A}{_B}+\frac{1}{2}\chi^{A}{_B}
+\frac{1}{2}x\alpha _{,x}\partial _{u}\chi^{A}{_B}
 +\frac{1}{2}x^{2}\left(3H^{A}{_B}
 -\frac{\chi ^{CD}\chi _{CD}}{4}\chi^{A}{_B}\right)\nonumber \\
 &  & {}+\frac{1}{4}x^{2}\alpha _{,x}\left(\chi _{BC}\partial _{u}\chi ^{AC}
 -\chi ^{AC}\partial _{u}\chi _{BC}
 +\chi ^{AD}{_{||DB}}-\chi _{BD}{^{||DA}}\right)+O(x^{3})\;,\nonumber
\end{eqnarray*}
\[
\Gtrzy ^{A}{_{BC}}={\Gamma(\hst) }^{A}{_{BC}}+O(x)\;.\]

\paragraph{Extrinsic curvature.
} The tensor field \( K_{ij} \) can be computed as \[
K_{ij}=\frac{1}{2N}(S_{i|j}+S_{j|i}-\partial _{u}\gmetric
_{ij})\;,\] where by \( A_{|i} \) we denote the covariant
derivative of a quantity \( A \) with respect to \( \gmetric
_{ij}. \) These covariant derivatives read:\begin{eqnarray*}
S_{x|x} & = & \frac{1}{x^{3}}\left[-1-\frac{x\alpha
_{,xx}}{2\alpha _{,x}}
+x\beta _{\uder x}-2\beta -\frac{1}{2}x^{2}\alpha _{,x}\right.\\
 &  & \left.{}+\frac{5}{2}x^{3}\alpha _{,x}M+x\alpha _{,x}\partial _{u}\beta +O(x^{4})
 \right]\;,
\end{eqnarray*}
\begin{eqnarray*}
S_{x|A} & = & \frac{\beta _{\uder A}}{x^{2}}-\frac{\alpha
_{,A}}{2x^{3}\alpha _{,x}}
-\frac{\alpha _{,xA}}{2x^{2}\alpha _{,x}}+\frac{1}{4x}\chi^{C}{_{A||C}}-N_{A}\\
 &  & {}-\frac{1}{32}(\chi ^{CD}\chi _{CD})_{||A}
 -\frac{1}{8}\chi _{AC}\chi^{CD}{_{||D}}+O(x)\;,
\end{eqnarray*}
 \begin{eqnarray*}
S_{A|x} & =& -\frac{\beta _{\uder A}}{x^{2}}-\frac{\alpha
_{,A}}{2x^{3}\alpha _{,x}}
-\frac{\alpha _{,xA}}{2x^{2}\alpha _{,x}}+\frac{1}{4x}\chi^{C}{_{A||C}}-3N_{A}\\
 &  & {}-\frac{3}{32}(\chi ^{CD}\chi _{CD})_{||A}-\frac{1}{8}\chi _{AC}\chi^{CD}{_{||D}}\\
 &  & {}+\frac{1}{2}\alpha _{,x}\partial _{u}(\chi^{C}{_{A||C}})+O(x)\;,
\end{eqnarray*}
\[
S_{A|B}=-\frac{1}{x^{2}}\left[\frac{\hst _{AB}}{2x\alpha _{,x}}
+\frac{\chi _{AB}}{4\alpha _{,x}}-\frac{1}{4}x\hst _{AB}
-\frac{1}{4}x\partial _{u}\chi _{AB}+O(x^{2})\right]\;.\]
Derivatives of \( \gmetric _{ij}\) with respect to \( u:\)\[
\partial _{u}\gmetric _{xx}=4x^{-2}\alpha _{,x}\partial _{u}\beta +O(x)\;,\]
\[
\partial _{u}\gmetric _{xA}=\frac{1}{2}\alpha _{,x}\partial _{u}(\chi^{C}{_{A||C}})+O(x)\;,\]
\[
\partial _{u}\gmetric _{AB}=x^{-1}\partial _{u}\chi _{AB}+O(1)\;.\]
Hence the extrinsic curvature:\begin{eqnarray*} K_{xx} & = &
\frac{\sqrt{2\alpha _{,x}}}{x^{2}} \left[-1-\frac{x\alpha
_{,xx}}{2\alpha _{,x}}+x\beta _{\uder x}-\beta
-\frac{1}{4}x^{2}\alpha _{,x}\right.\\
 &  &\left. {}+2x^{3}\alpha _{,x}M-x\alpha _{,x}\partial _{u}\beta +O(x^{4})\right]\;,
\end{eqnarray*}
\begin{eqnarray*}
K_{xA} & = & \frac{\sqrt{2\alpha
_{,x}}}{2}\left[\frac{1}{2}\chi^{C}{_{A||C}}
+\frac{\alpha _{,A}}{x^{2}\alpha _{,x}}-\frac{\alpha _{,xA}}{x\alpha _{,x}}-4xN_{A}
\right.\\
 &  &\left. {}-\frac{1}{8}x(\chi ^{CD}\chi _{CD})_{||A}
 -\frac{1}{4}x\chi _{AC}\chi^{CD}{_{||D}}+O(x^{2})\right.]\;,
\end{eqnarray*}
\[
K_{AB}=-\frac{\sqrt{2\alpha _{,x}}}{2x}\left[\frac{\hst
_{AB}}{x\alpha _{,x}} -\frac{\beta \hst _{AB}}{x\alpha
_{,x}}+\frac{\chi _{AB}}{2\alpha _{,x}} -\frac{3}{4}x\hst
_{AB}+\frac{1}{2}x\partial _{u}\chi _{AB}+O(x^{2})\right]\;.\] The
trace \( \trK=K^{i}{_i} \) can be easily calculated more precisely
as\[ \trK=\frac{1}{2N}[2S^{i}{_{|i}}-\gmetric ^{ij}\partial
_{u}\gmetric _{ij}]\;,\] or after multiplying by \( \sqrt{\det
\gmetric} \) and changing the covariant divergence to the ordinary
one:\[ \sqrt{\det \gmetric}\trK=\frac{1}{2N}\left[2(\sqrt{\det
\gmetric}S^{i})_{,i} -\frac{\partial _{u}(\det
\gmetric)}{\sqrt{\det \gmetric}}\right]\;.\] After some
calculations we get\begin{eqnarray*} \trK &
=-\frac{1}{\sqrt{2\alpha _{,x}}} & \left[3-3\beta -x\beta _{\uder
x}
+\frac{x\alpha _{,xx}}{2\alpha _{,x}}-\frac{3}{4}x^{2}\alpha _{,x}\right.\\
 &  & \left.{}-\frac{1}{2}x^{3}\alpha _{,x}\chi^{CD}{_{||CD}}
 +x\alpha _{,x}\beta _{,u}+O(x^{4})\right]\;.\nonumber
\end{eqnarray*}

\paragraph{ADM momenta.
} The ADM momenta can be expressed in terms of the extrinsic
curvature:\[ P^{k}{_l}=\gmetric ^{ki}(\gmetric
_{il}\trK-K_{il})\;.\] Substituting the previously calculated \(
\trK \) and \( K_{ij} \) we get\[
P^{x}{_x}=-\frac{1}{\sqrt{2\alpha _{,x}}}\left[2-2\beta
-\frac{3}{2}x^{2}\alpha _{,x}(1-2Mx) -\frac{1}{2}x^{3}\alpha
_{,x}\chi^{CD}{_{||CD}}+O(x^{4})\right]\;,\]
\begin{equation}
\label{PP0} P^{x}{_x}-\zP^{x}{_x}=\frac{-1}{\sqrt{2\alpha _{,x}}}
\left[-2\beta +3x^{3}\alpha _{,x}M-\frac{1}{2}x^{3}\alpha
_{,x}\chi^{CD}{_{||CD}}+O(x^{4})\right]\;,
\end{equation}
\begin{eqnarray} \nonumber
P^{A}{_B}&=& -\frac{1}{\sqrt{2\alpha _{,.x}}}\left[(2-2\beta -x\beta
_{\uder x} +\frac{x\alpha _{,xx}}{2\alpha
_{,x}})\delta^{A}{_B}+\frac{1}{2}x\chi^{A}{_B}
-\frac{1}{2}x^{2}\alpha _{,x}\partial
_{u}\chi^{A}{_B}\right]\\ && +O(x^{3})\;,\nonumber \end{eqnarray}
\[
P^{A}{_x}=-\frac{1}{2}x^{2}\sqrt{2\alpha
_{,x}}\chi^{AC}{_{||C}}+O(x^{3})\;,\]
\begin{eqnarray} \nonumber
P^{x}{_A}&=& \frac{-x^{3}}{2\sqrt{2\alpha _{,x}}}
\left[\frac{\chi^{C}{_{A||C}}}{x}-\frac{\alpha _{,xA}}{x^{2}\alpha
_{,x}} -6N_{A}-\frac{3(\chi ^{CD}\chi _{CD})_{||A}}{16}
-\frac{1}{2}\chi _{AC}\chi^{CD}{_{||D}}\right]
\\ && +O(x^4) \;, \nonumber \end{eqnarray}
\begin{eqnarray} \nonumber
 P^{x}{_A}-\zP^{x}{_A} & = &\frac{-x^{3}}{2\sqrt{2\alpha
_{,x}}}\left[\frac{\chi^{C}{_{A||C}}}{x}-6N_{A} -\frac{3(\chi
^{CD}\chi _{CD})_{||A}}{16} -\frac{1}{2}\chi _{AC}\chi
^{CD}{_{||D}}\right] \\ &&  +O(x^4) \;.
\label{PPA}
\end{eqnarray}

\paragraph{The second fundamental form \( k_{AB} \).
} The extrinsic curvature of the leaves of the ''2+1 foliation''
can be computed from the formula\[
k_{AB}=\frac{\Gtrzy^{x}{_{AB}}}{\sqrt{\gmetric ^{xx}}}\;.\]
Hence\begin{eqnarray*} k_{AB} & = & \frac{\sqrt{2\alpha
_{,x}}}{2x}\left[\frac{1}{x\alpha _{,x}}\hst _{AB}
+\frac{1}{2\alpha _{,x}}\chi _{AB}\right.\\
 &  &\left. {}+\frac{1}{4}x\hst _{AB}-\frac{1}{2}x\partial _{u}\chi _{AB}
 -\frac{\beta }{x\alpha _{,x}}\hst _{AB}+O(x^{2})\right]\;.\nonumber
\end{eqnarray*}
We need a more accurate expansion of the trace \( k=\gtw
^{AB}k_{AB}. \) The formula\[ k_{AB}=\frac{\Gamma
^{x}{_{AB}}}{\sqrt{\gmetric ^{xx}}} -\frac{\fourg^{x\omega }\Gamma
^{\omega }{_{AB}}}{\fourg^{\omega \omega }\sqrt{\gmetric
^{xx}}}\;.\] can be used. It is convenient to calculate \( \gtw
^{AB}\Gamma ^{x}{_{AB}} \) and \( \gtw ^{AB}\Gamma ^{\omega
}{_{AB}} \) using the expressions for the Christoffel symbols
given in \cite[Appendix C]{CJK}:\[ \Gamma
^{\omega}{_{AB}}=x^{-1}\mathrm{e}^{-2\beta
}(h_{AB}-\frac{1}{2}xh_{AB\uder x})-\alpha _{,x}\Gamma
^{x}{_{AB}}+O(x^{3})\;,\]
\[
\Gamma ^{x}{_{AB}}=-\frac{1}{2}\mathrm{e}^{-2\beta }(2\Der
(h)_{(A}U_{B)}+\partial _{u}h_{AB}-2Vx^{2}h_{AB}+Vx^{3}h_{AB\uder
x})\;,\] where \( \Der (h)_{A} \) is a covariant derivative with
respect to \( h_{AB} \) (we use the \( (u,x,x^{A}) \) coordinate
system and all the differentiations with respect to \( x, \) \(
x^{A} \) are at constant \( u \)). Hence:\[ \gtw ^{AB}\Gamma
^{\omega }{_{AB}} =2x-4\beta x+x^{2}\alpha
_{,x}(U^{A}{_{||A}}-2Vx^{2})+O(x^{5})\;,\]
\[
\gtw ^{AB}\Gamma ^{x}{_{AB}}=
-x^{2}[U^{A}{_{||A}}-2Vx^{2}+O(x^{3})]\;.\] After substitution of
relevant asymptotic expansions:\[ k=\frac{1}{\sqrt{\gmetric
^{xx}}} \left[\frac{1}{4}x^{4}\chi^{CD}{_{||CD}}
+\frac{1}{2}x^{3}(1-2Mx)+\frac{x}{\alpha _{,x}} -\frac{2\beta
x}{\alpha _{,x}}+O(x^{5})\right]\;,\]
\[
k=\sqrt{2\alpha _{,x}}\left[\frac{1}{4}x^{3}\chi^{CD}{_{||CD}}
+\frac{1}{4}x^{2}(1-2Mx)+\frac{1}{\alpha _{,x}} -\frac{\beta
}{\alpha _{,x}}+O(x^{4})\right]\;,\] \be k-\zk  =\sqrt{2\alpha
_{,x}}\left[\frac{1}{4}x^{3}\chi^{CD}{_{||CD}}
-\frac{1}{2}x^{3}M-\frac{\beta }{\alpha
_{,x}}+O(x^{4})\right]\;.\label{kk0} \ee

\subsection{The polyhomogenous case}
\label{app:obiekty3dlog} We give only the most important
intermediate results which differ from the power-series case:
\[
\gmetric
^{xB}=-\frac{1}{4}x^{4}\chi^{AC}{_{||C}}+\frac{1}{2}x^{5}W^{A}+O(x^{5})\;,\]
\[
S_{A}=\frac{1}{2}\chi^{C}{_{A||C}}-xW_{A}+O(x)\;,\]
\[
K_{xA}=\sqrt{2\alpha _{,x}}[\frac{1}{4}\chi^{C}{_{A||C}}-xW_{A}
-\frac{1}{2}x^{2}W_{A\uder x}+O(x)]\;,\]
\[
P^{x}{_A}=-\frac{x^{2}}{2\sqrt{2\alpha
_{,x}}}[\chi^{C}{_{A||C}}-3xW_{A}-x^{2}W_{A\uder x}+O(x)]\;.\]

\section{Decomposition of Poincar\'e group vectors into tangential and normal parts}

\label{app:killingi}The generators of Poincar\'e group are given
after \cite{CJK}:\[ X_{time}=\partial _{\omega }=\partial
_{u}\;,\]
\[
X_{rot}=-\varepsilon ^{AB}\alpha _{,A}v_{,B}\partial _{\omega
}+\varepsilon ^{AB}v_{,B}\partial _{A}\;,\]
\[
X_{trans}=(-v-x\alpha ^{,A}v_{,A}+x^{2}v\alpha _{,x})\partial
_{\omega }-x^{2}v\partial _{x} +xv^{,A}\partial _{A}\;,\]
\begin{eqnarray*}
X_{boost} & = & [xv((\alpha +\omega )x+1)\alpha _{,x}-(\alpha +\omega )v\\
 & - & \alpha ^{,A}v_{,A}((\alpha +\omega )x+1)]\partial _{\omega }\nonumber \\
 & - & xv[(\alpha +\omega )x+1]\partial _{x}+v^{,A}[(\alpha +\omega )x+1]\partial _{A}\;.\nonumber
\end{eqnarray*}
The tensor \( \varepsilon ^{AB} \) is defined as \[ \varepsilon
^{AB}=\frac{1}{\sqrt{\hst }}\{A,B\}\;,\] where \(
\{1,2\}=-\{2,1\}=1 \) and \( \{1,1\}=\{2,2\}=0. \) By \( v \) we
denote a function on the sphere which is a combination of $\ell=1$
spherical harmonics. If we consider embedding of the sphere into
\( \real ^{3}, \) then
\[ v(x^{A})=v^{i}\frac{x_{i}}{r}\] and we get a bijection between
functions \( v \) and vectors \( (v^{i})\in \real ^{3}. \)

Let us now decompose a vector field \( X \) into parts tangent and
normal to  \( \hyp : \)\[ X=Y+Vn\;.\] \( Y \) is a vector tangent
to \( \hyp \  \) and \( n \) is a unit (\( n^{2}=-1 \)),
future-directed normal vector. Setting $$ \tau =2\alpha _{,x}-\hst
^{AB}\alpha _{,A}\alpha _{,B} \;,$$ we have the following
decomposition of respective vectors : \[
V_{time}=\frac{1}{x\sqrt{\tau }}\left(1-\frac{x\chi ^{CD}\alpha
_{,C}\alpha _{,D}}{2\tau }+O(x^{2})\right)\;,\]
\[
Y^{x}_{time}=\frac{1}{\tau }\left(1-\frac{x\chi ^{CD}\alpha
_{,C}\alpha _{,D}}{\tau }+O(x^{2})\right)\;,\]
\[
Y^{A}_{time}=-\frac{1}{\tau }\left(a^{,A}-x\chi ^{AC}\alpha _{,C}
-\frac{x\chi ^{CD}\alpha _{,C}\alpha _{,D}\alpha ^{,A}}{\tau
}+O(x^{2})\right)\;,\]
\[
V_{rot}=-\varepsilon ^{AB}\alpha _{,A}v_{,B}\cdot
\frac{1}{x\sqrt{\tau }} \left(1-\frac{x\chi ^{CD}\alpha
_{,C}\alpha _{,D}}{2\tau }+O(x^{2})\right)\;,\]
\[
Y^{x}_{rot}=-\varepsilon ^{AB}\alpha _{,A}v_{,B}\cdot
\frac{1}{\tau } \left(1-\frac{x\chi ^{CD}\alpha _{,C}\alpha
_{,D}}{\tau }+O(x^{2})\right)\;,\]
\begin{eqnarray*}
Y^{A}_{rot} & = & \varepsilon ^{AB}v_{,B}\\
 && {}+  \varepsilon ^{CB}\alpha _{,C}v_{,B}\cdot \frac{1}{\tau }
 \left(\alpha ^{,A}-x\chi ^{AC}\alpha _{,C}
 -\frac{x\chi ^{CD}\alpha _{,C}\alpha _{,D}\alpha ^{,A}}{\tau }+O(x^{2})\right)\;,\nonumber
\end{eqnarray*}
\[
V_{trans}=\frac{1}{x\sqrt{\tau }}\left(-v-x\alpha ^{,A}v_{,A}
+v\frac{x\chi ^{CD}\alpha _{,C}\alpha _{,D}}{2\tau
}+O(x^{2})\right)\;,\]
\[
Y^{x}_{trans}=\frac{1}{\tau }\left(-v-x\alpha ^{,A}v_{,A}
+v\frac{x\chi ^{CD}\alpha _{,C}\alpha _{,D}}{\tau
}+O(x^{2})\right)\;,\]
\begin{eqnarray*}
Y_{trans}^{A} & = & xv^{,A}\\
 && \hspace*{-1cm} {}-  \frac{1}{\tau }\left(-v\alpha ^{,A}+vx\chi ^{AC}\alpha _{,C}
 +v\frac{x\chi ^{CD}\alpha _{,C}\alpha _{,D}}{\tau }\alpha ^{,A}
 -x\alpha ^{,C}v_{,C}\alpha ^{,A}+O(x^{2})\right)\;,\nonumber
\end{eqnarray*}
\begin{eqnarray*}
V_{boost} & = & \frac{1}{x\sqrt{\tau }}\Big[-(\omega +\alpha )v
+xv\alpha _{,x}-v^{,C}\alpha _{,C}(\omega x+\alpha x+1)\\
 && {}+  \frac{x\chi ^{CD}\alpha _{,C}\alpha _{,D}}{2\tau }((\omega +\alpha )v+
 v^{,C}\alpha _{,C})+O(x^{2})\Big]\;,\nonumber
\end{eqnarray*}
\begin{eqnarray*}
Y^{x}_{boost} & = & -xv+\frac{1}{\tau }\Big[-(\omega +\alpha
)v+xv\alpha _{,x}-
v^{,C}\alpha _{,C}(\omega x+\alpha x+1)\\
 && {} +  \frac{x\chi ^{CD}\alpha _{,C}\alpha _{,D}}{\tau }\big((\omega +\alpha )v+
 v^{,C}\alpha _{,C}\big)+O(x^{2})\Big]\;,\nonumber
\end{eqnarray*}
\begin{eqnarray*}
Y^{A}_{boost} & = & v^{,A}(\omega x+\alpha x+1)\\
 && {}+  \frac{1}{\tau }\Big[(\omega +\alpha )v\alpha ^{,A}-xv\alpha _{,x}\alpha ^{,A}
 +v^{,C}\alpha _{,C}\alpha ^{,A}(\omega x+\alpha x+1)\nonumber \\
 && {}-  x\big(\frac{\chi ^{CD}\alpha _{,C}\alpha _{,D}\alpha ^{,A}}{\tau }
 +\chi ^{AC}\alpha _{,C}\big)\big((\omega +\alpha )v+v^{,C}\alpha _{,C}\big)+O(x^{2})\Big]\;.\nonumber
\end{eqnarray*}


\begin{thebibliography}{10}
\bibitem{ChAIHP}
\bysame, \emph{On the relation between the {Einstein} and the
{Komar}
  expressions for the energy of the gravitational field}, Ann. Inst. H.
  Poincar\'e \textbf{42} (1985), 267--282.
\end{thebibliography}

\def\cprime{$'$}
\providecommand{\bysame}{\leavevmode\hbox
to3em{\hrulefill}\thinspace}
\providecommand{\MR}{\relax\ifhmode\unskip\space\fi MR }
\providecommand{\MRhref}[2]{%
  \href{http://www.ams.org/mathscinet-getitem?mr=#1}{#2}
} \providecommand{\href}[2]{#2}

\end{document}